\documentclass[journal,twocolumn,10pt,twoside]{IEEEtran}
\usepackage{graphicx,cite,epsfig,amssymb,amsmath,amsfonts,bbm}
\usepackage{pifont}
\usepackage{stmaryrd}
\usepackage{bbding}
\usepackage{subfigure}
\usepackage{setspace}
\usepackage{color}
\usepackage{slashbox}
\usepackage{algorithm,algorithmic}
\usepackage{bm}
\usepackage{subfigure}
\newtheorem{lemma}{Lemma}
\allowdisplaybreaks[4]

\begin{document}

\title{Design and Analysis of Massive Uncoupled Unsourced Random Access with Bayesian Joint Decoding}
\author{\normalsize
Feiyan Tian, Xiaoming Chen, Yong Liang Guan, and Chau Yuen

\thanks{Feiyan Tian and Xiaoming Chen are with the College of Information Science and Electronic Engineering, Zhejiang University, Hangzhou 310027, China (e-mail: \{tian\_feiyan, chen\_xiaoming\}@zju.edu.cn).
Yong Liang Guan and Chau Yuen are with the school of Electrical and Electronic Engineering, Nanyang Technological University, Singapore 639798 (e-mail: \{eylguan, chau.yuen\}@ntu.edu.sg).}}\maketitle

\begin{abstract}
In this paper, we investigate unsourced random access for massive machine-type communications (mMTC) in the sixth-generation (6G) wireless networks.
Firstly, we establish a high-efficiency uncoupled framework for massive unsourced random accesss without extra parity check bits.
Then, we design a low-complexity Bayesian joint decoding algorithm, including codeword detection and stitching. In particular, we present a Bayesian codeword detection approach by exploiting Bayes-optimal divergence-free orthogonal approximate message passing in the case of unknown priors. The output long-term channel statistic information is well leveraged to stitch codewords for recovering the original message. Thus, the spectral efficiency is improved by avoiding the use of parity bits. Moreover, we analyze the performance of the proposed Bayesian joint decoding-based massive uncoupled unsourced random access scheme in terms of computational complexity and error probability of decoding. Furthermore, by asymptotic analysis, we obtain some useful insights for the design of massive unsourced random access. Finally, extensive simulation results confirm the effectiveness of the proposed scheme in 6G wireless networks.
\end{abstract}


\section{Introduction}
The sixth-generation (6G) wireless networks are expected to provide various machine-type communication services for Internet-of-Things (IoT) \cite{5G1,5G2,GFR1}. Generally, machine-type communications in 6G wireless networks have the demands of massive connectivity and small payload. It is predicted that by 2030, the number of IoT devices will approach hundreds of billions and the length of a message is usually tens of bits. For such massive machine-type communications (mMTC), applying traditional grant-based random access schemes may lead to exceedingly high access latency and prohibitive signaling overhead \cite{GF0}. To address these challenging issues, it is imperative to introduce grant-free random access schemes for supporting massive access according to the characteristics of mMTC in 6G wireless networks \cite{GF1,GF2,GFR2}.

As a new grant-free random access scheme, unsourced random access overcomes the disadvantages of sourced grant-free random access, i.e., the leading process of acquisition of the device activity information and channel state information has been cancelled, such that the end-to-end delay is decreased and the spectral efficiency is increased \cite{UN1,UN2}.
To be specific, in the unsourced random access protocol, the base station (BS) only focuses on the recovery of transmitted messages, but is not concerned about the identities and channel states of active devices which sent the messages \cite{UN8,UN9,UN10}. In this context, sending unique long pilots in advance is not necessary for device detection and channel estimation, which reduces the consumption of wireless resources significantly in the case of massive connectivity. In fact, the biggest superiority of unsourced random access over sourced random access is that the millions of individual codebooks of devices are avoided \cite{UN1}. Hence, all devices can share the same codebook in unsourced random access. The message of each active device is transmitted over uplink channel after mapping to a codeword from this common codebook. Subsequently, the BS performs codeword activity detection and recovers the list of transmitted codewords.
Note that a codebook needs to contain at least $2^b$ codewords to represent $b$-bit message. As the length of the message $b$ increases, the size of the codebook $2^b$ grows exponentially, resulting in unbearable computational complexity for codeword activity detection even with short message of tens of bits. In order to reduce the computational complexity for codeword activity detection, a divide-and-send approach is introduced and each segmentation is independently sent based on a small codebook \cite{UN01}.

Recently, the authors of \cite{UN2} proposed a low-complexity scheme for unsourced random access based on a coupled compressed sensing problem. In their scheme, the message transmission slot is divided into several sub-slots and the message fragments are transmitted by each active device across these sub-slots. Specifically, at the sender, the message is first divided into multiple fragments by a outer encoder with redundancy attached to form fixed-length sub-blocks. Then an inner encoder maps each sub-block to a codeword via a common codebook matrix. At the BS, the inner decoder identifies which codewords have been transmitted and a tree-based outer decoder connects the decoded sub-blocks to recover original long messages according to the redundant parity.
Based on the proposed framework in \cite{UN2}, the authors of \cite{UN3} extended the system model from a single-antenna receiver to a large-scale receive antenna array, which widens application scenarios of coupled unsourced random access. In \cite{UN4}, approximate meessage passing (AMP) was exploited as an inner decoder to realize codeword activity detection. Additionally, codeword activity detection of unsourced random access in the inner decoder was formulated as a non-Bayesian maximum likelihood (ML) problem in \cite{UN5}. It is proved that this scheme has better stability over the Bayesian AMP schemes. In \cite{UN6}, a beam-space tree decoder was proposed to improve the decoding performance by exploiting the beam division property.
On the other hand, the authors of \cite{UN7} improved the unsourced random access framework in \cite{UN2} by letting the inner and the outer decoder cooperate and passing information back and forth to enhance the performance of coded compressed sensing.

Meanwhile, in \cite{UN11}, the message of each active device was just divided into two parts. The first part is to determine the transmission rule and the second part is encoded by low density parity check (LDPC) code. The involved sparse joint Tanner graph can provide some improvements in performance.
Similarly, the scheme that some message bits were mapped to pilot and spreading sequences and the other bits were processed by polar code and QPSK modulation was studied in \cite{UN17}, providing good solutions for supporting a large number of active UEs with finite block length.
In addition, the authors of \cite{UN18} also proposed an elegant semi-blind detection framework based on bilinear generalized approximate message passing algorithm, which can support both sourced and unsourced random access.

Although the redundancy introduced by coupled unsourced random access is limited compared with the pilot sequences in the sourced random access, in order to ensure the correctness and uniqueness of output of the outer decoder, these additional redundant parity bits usually occupy half or more of each sub-block, resulting in low spectral efficiency of the system, which is intolerable under the shortage of wireless resources.
In this context, a distinct uncoupled compressed sensing-based unsourced random access scheme was proposed in \cite{UN12}. At the sender, the message is not attached to the check bits and the tree-based outer decoder is changed into a clustering decoder, where the disordered sub-blocks are connected by leveraging the inherent correlations of instantaneous channels of the same active device across sub-slots. Further, the authors of \cite{UN13} studied the unsourced random access scheme by exploiting the angular domain sparsity of channel based on this uncoupled framework, which has a better error performance with high spectral efficiency.

As a matter of fact, codeword connection by clustering instantaneous channels may be prone to errors because the channel state is not invariable in a long time slot. Intuitively, it is more reasonable to assume that statistic information of channel is constant in a certain long time slot. Hence, it makes sense to exploit the statistics of channel to implement the codeword stitching.
Besides, the scheme in \cite{UN13} is limited to 3D channel modeling in angle domain, leading to higher implementation complexity. In other words, existing methods trade complexity for the decoding performance of unsourced random access.
In addition, to the best of authors' knowledge, the error performance analysis of massive unsourced random access is still an open issue.

To improve the performance of unsourced random access schemes, we aim to provide a high-efficiency and low-complexity solution for uncoupled unsourced random access in 6G wireless networks. The contributions of this paper are as follows:

\begin{enumerate}
\item We provide a high-efficiency unsourced random access scheme inspired by the recent uncoupled framework in which parity bits are avoided but channel characteristics are exploited for codeword concatenation.

\item We design a low-complexity Bayesian joint decoding algorithm, which implements codeword detection in the case of unknown priors and codeword stitching with the assistance of channel statistics.

\item We analyze the overall performance of the proposed Bayesian joint decoding-based uncoupled unsourced random access scheme, and confirm its low complexity and high reliability.

\item We obtain some useful insights by asymptotic analysis and prove that the error probability of codeword detection tends to zero by increasing the number of BS antennas and transmit power.

\end{enumerate}

The rest of this paper is organized as follows: Section II introduces the uncoupled unsourced random access-based 6G mMTC model. Section III designs a Bayesian joint decoding algorithm for the proposed massive uncoupled unsourced random access scheme. Section IV analyzes the convergence, complexity and error performance of the proposed decoding algorithm and provides some useful insights via asymptotic analysis.
Then, simulation results are given in Section V to evaluate the effectiveness of the proposed scheme. Finally, Section VI concludes the paper.

\emph{Notations}: Bold upper (lower) letters denote matrices (column vectors), $(\cdot)^T$ denotes transpose, $(\cdot)^H$ denotes conjugate transpose, $\mathbb{C}^{a \times b}$ denotes a complex matrix or vector of dimension $a \times b$, $\mathcal{CN}(\textbf{x},\textbf{Y})$ denotes the complex Gaussian distribution of a vector with mean $\textbf{x}$ and covariance $\textbf{Y}$, $\textmd{Pr}(\cdot)$ denotes the probability of an event, $\mathbb{E}\{\cdot\}$ denotes expectation, $\exp(\cdot)$ denotes the exponent, $[\textbf{X}]_{a,b}$, $[\textbf{X}]_{a,:}$ and $[\textbf{X}]_{:,b}$ denote the $(a,b)$-th element, $a$-th row and $b$-th column of matrix $\textbf{X}$, respectively, $\|\cdot\|_2$ and $\|\cdot\|_F$ denote the 2-norm of a vector and Frobenius norm of a matrix respectively, and $\textmd{tr}(\cdot)$ denotes the trace of a matrix. $[\mathcal{L}]_{k}$ denotes the $k$-th element of set $\mathcal{L}$.

\section{System Model}
Consider an unsourced random access protocol-based single-cell 6G mMTC system, where a BS equipped with $M$ antennas serves $K_{\textmd{tot}}$ single-antenna IoT user equipments (UEs) over the same time-frequency resource block. Due to the sporadic traffic of IoT applications, only a small set of $K_a(K_a \ll K_{\textrm{tot}})$ UEs, denoted by $\mathcal{K}_a$, are active in a certain time slot. Active UE $k$ transmits a binary message $\bm{m}_k$ of $b$ bits to the BS and the transmitted message set is represented by $\mathcal{L}=\{\bm{m}_k : k \in \mathcal{K}_a\}$.
To reduce the complexity of receiver, the $b$-bit message $\bm{m}_k$ of the $k$-th $(k \in \mathcal{K}_a)$ active UE is divided into $L$  short sub-blocks of length $J$ with $b=LJ$, such that a small codebook can be employed. Meanwhile, a time slot is partitioned into $L$ sub-slots of duration $n_0$ symbols each.
According to the principle of unsourced random access, all UEs share the same codebook. Each active UE maps its sub-blocks to codewords based on a common codeword selection scheme.
Specifically, let $\textbf{C}=[\bm{c}_1,...,\bm{c}_{2^J}]\in \mathbb{C}^{n_0\times2^J}$ be the codebook with each column $\{\bm{c}_i\in \mathbb{C}^{n_0\times1},\ i\in [1:2^J]\}$ representing a codeword with unit norm $\|\bm{c}_i\|^2=1$. In the $l$-th ($l\in [1,L]$) sub-slot, the $l$-th sub-block of the $k$-th ($k\in \mathcal{K}_a$) active UE is mapped to integer $i_{k,l}$ and the $i_{k,l}$-th codeword $\bm{c}_{i_{k,l}}$ will be transmitted. For a $J$-bit binary sub-block, there are $2^J$ possible combinations, thus $i_{k,l}\in [1,2^J]$.
Herein, the split of message and mapping of codeword at the active UE terminal are called encoding.


After mapping, all active UEs synchronously transmit their codewords to the BS over $L$ sub-slots in sequence. Hence, the received signal $\textbf{Y}_l \in \mathbb{C}^{n_0 \times M}$ at the BS in the $l$-th sub-slot can be expressed as
\begin{eqnarray}\label{model}
\textbf{Y}_l=\sum\limits_{k\in \mathcal{K}_a}\bm{c}_{i_{k,l}}\bm{h}^T_{k,l}+\textbf{Z}_l
=\textbf{C}\boldsymbol{\Delta}_l\textbf{H}_l+\textbf{Z}_l
=\textbf{C}\textbf{X}_l+\textbf{Z}_l,
\end{eqnarray}
where $\bm{c}_{i_{k,l}}$ denotes the codeword sent by the $k$-th active UE in the $l$-th sub-slot. $\bm{h}_{k,l} \in \mathbb{C}^{M \times 1}$ is the channel vector from the UE $k$ to the BS in the $l$-th sub-slot. It follows Rayleigh block fading across sub-slots, i.e., $\bm{h}_{k,l}\sim \mathcal{CN}(\bm{0},\tilde{g}_k\mathbf{I})$ with $\tilde{g}_k$ being the large-scale fading coefficient which are unknown at the BS. $\textbf{Z}_l$ represents the additive white Gaussian noise (AWGN) with zero mean and variance $\sigma^2=N_0/(n_0P_t)$, where $N_0$ is noise power and $P_t$ is the per-symbol maximum transmit power.
By arrangement, we let $\textbf{H}_l=[\bm{h}_{1,l},\bm{h}_{2,l},...,\bm{h}_{{K_{\textrm{tot}}},l}]^T$. $\boldsymbol{\Delta}_l \in \{0,1\}^{2^J \times K_{\textrm{tot}}}$ is a binary codeword activity indicator matrix. It has 1 elements in the $i_{k,l}$-th row and the $k$-th ($k\in\mathcal{K}_a$) column, and 0 entries in the rest.
With the received signals, the BS detects active slot-wise codewords $\bm{c}_{i_{k,l}}$ and stitches the codewords across sub-slots to recover messages $\bm{m}_k$. In the following section, we design a decoding algorithm for the BS to achieve this goal.

\section{Design of Bayesian Joint Decoding}
In this section, we design a Bayesian joint decoding algorithm to recover the original messages from the received signals at the BS. Specifically, the decoder detects the transmitted codewords of every sub-slot based on noisy observation $\textbf{Y}_l$ and common codebook $\textbf{C}$ without prior information including the number of active UEs $K_a$, channel fading coefficient $\tilde{g}_k$, and noise variance $\sigma^2$. Then, the acquired codewords are unmapped to sub-blocks and stitched together to obtain original messages. The details of the Bayesian joint decoding algorithm are provided below.

\subsection{The preparation of decoding}
\emph{Assumption:} In this paper, to facilitate decoding design and performance analysis, it is assumed that the transmitted codewords within a certain sub-slot do not collide with each other due to suitable parameter settings\footnotemark[1]$^{,2}$.

\footnotetext[1]{In general, the $2^J$ codewords in the common codebook are selected by $K_a$ active UEs independently across $L$ sub-slots. The codeword collision probability can be computed as $\textmd{Pr}(\textmd{collision})=1-((1-2^{-J})^{(K_a-1)})^L$. For convenience of decoding design and performance analysis, the ideal situation without collision under suitable parameters is considered.

$^2$In the case of codeword collision, the following collision intervention mechanism before codeword stitching is given for reference \cite{retrans}. When collisions occur in sub-slot $l$, the number of non-zero rows of the codeword detection output $\hat{\textbf{X}}_l$ is less than $K_a$ (the estimated $K_a$ in the former sub-slot without collisions can be used) and the BS can only recover the superimposed channels of collided codewords, resulting in the failure of codeword concatenation. At this time, the BS judges which codewords are sent by multiple UEs via energy detection and then feeds the indices of repeatedly transmitted codewords back to all UEs. The UEs who find they are in a collision slide their sub-block $l$ window of length $J$ bits forward with sliding length $0<L_{\rm slide}<J$, such that the new sequences can be used to map different codewords. Note that these new codewords are retransmitted and only used to detect the channels of collided UEs, where the channels no longer overlap and can be used for the stitching of original codewords.}

\begin{figure}[h] \centering
\includegraphics [width=0.5\textwidth] {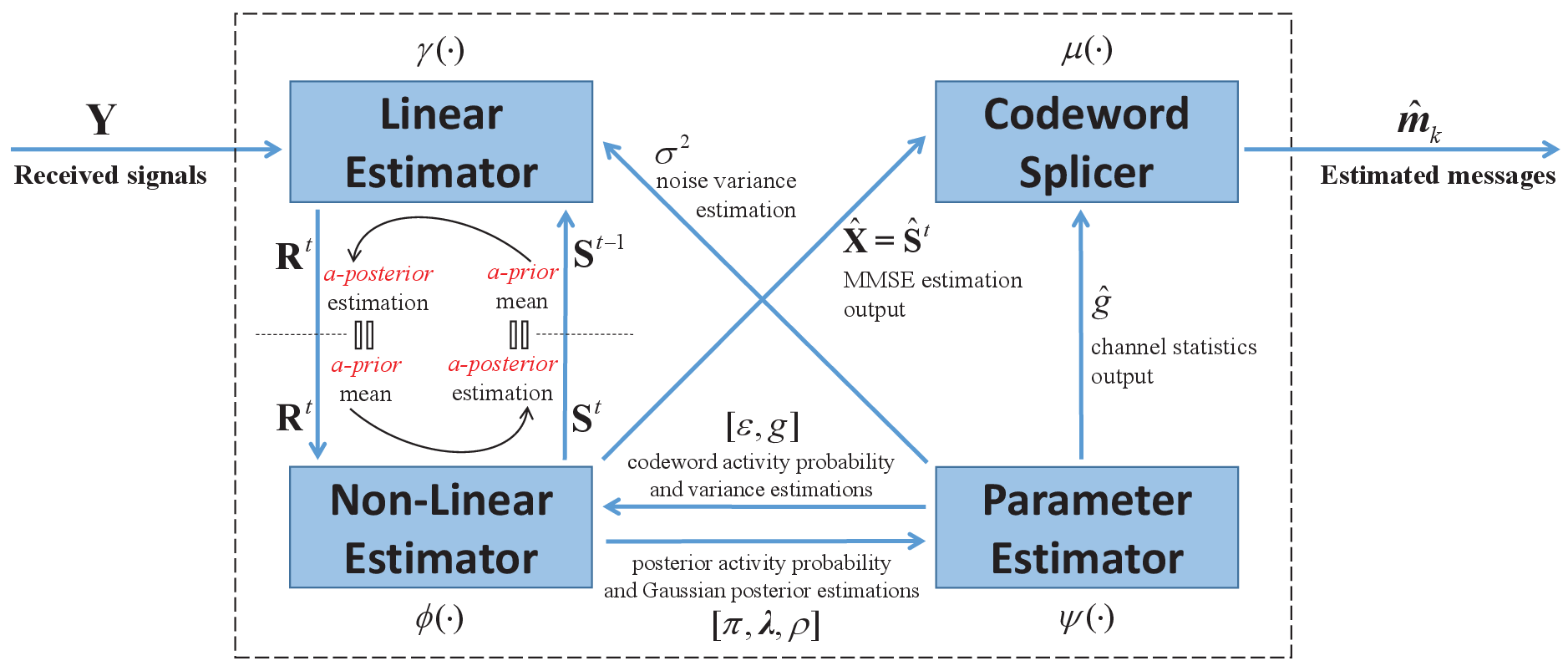}
\caption {The flow chart of Bayesian joint decoding. The decoder consists of four local modules, these modules work together to recover the original messages $\hat{\bm{m}}_k$ sent from multiple active UEs according to the noisy received signals $\textbf{Y}$. In particular, the OAMP detector including linear estimator $\gamma(\cdot)$ and non-linear estimator $\phi(\cdot)$ aims to detect the codeword state matrix $\textbf{X}$. The iteration between $\textbf{R}$ and $\textbf{S}$ has been defined in \eqref{Ite_LE} and \eqref{Ite_NLE}. By leveraging the MMSE estimations $[\pi, \boldsymbol{\lambda}, \rho]$ of OAMP detector, parameter estimator $\psi(\cdot)$ estimates the unknown system parameters $[\sigma^2,\varepsilon,g]$ and then feeds them back. With the estimated codeword state matrix $\hat{\textbf{X}}$ and channel statistics $\hat{g}$, the original messages $\hat{\bm{m}}_k$ can be reconstructed in codeword splicer $\mu(\cdot)$.}\label{Fig2}
\end{figure}

Under this assumption, the following approximate distribution can be adopted in decoding.
For the matrix $\textbf{X}_l\in \mathbb{C}^{2^J \times M}=\boldsymbol{\Delta}_l\textbf{H}_l$, its $j$-th row follows a Bernoulli-Gaussian distribution, i.e., $\forall j\in[1,2^J]$
\begin{equation}
\bm{x}_{j,l}\sim\left\{
\begin{array}{ll}
\mathcal{CN}(\bm{0},g_{j,l}\textbf{I}),&\textmd{probability}=\varepsilon_j\\
\textbf{0},&\textmd{probability}=1-\varepsilon_j\end{array}\right.,
\end{equation}
where $\varepsilon_j=1-(1-1/2^J)^{K_a}\approx K_a/2^J$ denotes the non-zero probability of $\bm{x}_{j,l}$, i.e., codeword activity probability. $g_{j,l}=\sum\nolimits_{k}\delta^l_{j,k}\tilde{g}_k$ represents the codeword variance with $\delta^l_{j,k}$ being the $(j,k)$-th element of  $\boldsymbol{\Delta}_l$. 
For notational simplicity, we omit the sub-slot index $l$ since the codewords transmission and detection are identical in each sub-slot, and denote the above distribution as
\begin{align}\label{distribution}
    P_{X}(\bm{x}_j;\varepsilon_j,g_j)=(1-\varepsilon_j)\delta_0 + \varepsilon_j\mathcal{CN}(\bm{x}_j;\bm{0},g_j\textbf{I}),
\end{align}
where $\delta_0$ is the dirac Delta at zero.
In this context, $\textbf{X}$ can be called a non-binary codeword state matrix and the row non-zero probability is $\varepsilon_j$. Due to the sporadic activation of UEs, matrix $\textbf{X}$ is row-wise sparse. Moreover, it is seen that the dimensions of codeword state matrix are independent of the total number of potential UEs $K_{\textmd{tot}}$.
Based on this model setting, the proposed Bayesian joint decoding algorithm containing codeword detection and stitching is implemented as illustrated in Fig. \ref{Fig2}.

\subsection{The detection of codeword state matrix}
Intuitively, the recovery of lists of transmitted codewords is equivalent to the reconstruction of codeword state matrix $\textbf{X}$, which is a compressed sensing problem because $\textbf{X}$ is row-wise sparse. Further, this is also a multiple measurement vectors (MMV) setup due to multiple BS antennas \cite{MMV}. By leveraging the sparsity structure of $\textbf{X}$ as prior, in this part, we exploit the Bayes-optimal and divergence-free orthogonal approximate message passing (OAMP) method to detect the codeword state matrix \cite{OAMP,VAMP,MMVOAMP}. Meanwhile, since the BS has no knowledge of codeword activity probability $\varepsilon_j$, codeword variance $g_j$ and noise variance $\sigma^2$, an expectation maximization (EM) algorithm is adopted to estimate them \cite{EM}.


Generally, we aim to obtain the minimum mean square error (MMSE) estimation of $\textbf{X}$ based on the received signals \eqref{model} and sparsity structure \eqref{distribution}. Yet, it is not trivial to address these two constraints jointly. To this end, the proposed OAMP detector works in an iterative way. Specifically, two local modules, i.e., a linear estimator $\gamma(\cdot)$ and a non-linear estimator $\phi(\cdot)$, process linear constraint \eqref{model} and non-linear constraint \eqref{distribution} separately and run iteratively to obtain the final results. Define the iteration between the $\gamma(\cdot)$ and $\phi(\cdot)$ as:
\begin{align}
\textbf{R}^t&=\gamma(\textbf{S}^{t-1}),\label{Ite_LE}\\
\textbf{S}^t&=\phi(\textbf{R}^{t})\label{Ite_NLE},
\end{align}
where $t$ is the iteration index, $\textbf{R}^t$ in \eqref{Ite_LE} (or $\textbf{S}^t$ in \eqref{Ite_NLE}) is the \emph{a-posterior} estimation of $\textbf{X}$ generated by estimator $\gamma(\cdot)$ (or $\phi(\cdot)$) and the \emph{a-prior} mean $\textbf{S}^{t-1}$ of $\textbf{X}$ in \eqref{Ite_LE} (or $\textbf{R}^t$ in \eqref{Ite_NLE}).

Now let's decompose the MMV problem caused by multiple-antenna deployment to simplify OAMP iterations. For linear constraint \eqref{model}, since the correlation between different antennas (i.e., different columns of $\textbf{X}$) is not considered in our paper, matrix operation $\gamma(\textbf{S}^{t-1})$ is equivalent to column-vector operation $\gamma([\textbf{S}^{t-1}]_{:,m}),m\in[1,M]$. For non-linear constraint \eqref{distribution}, because $\textbf{X}$ is row-sparse and channels between different codewords (i.e., different rows of $\textbf{X}$) are independent, matrix operation $\phi(\textbf{R}^t)$ is equivalent to row-vector operation $\phi([\textbf{R}^t]_{j,:}),j\in[1,2^J]$.

First, we consider the linear estimator $\gamma([\textbf{S}^{t-1}]_{:,m})$ based on the received column-vector signals $[\textbf{Y}]_{:,m}=\textbf{C}[\textbf{X}]_{:,m}+[\textbf{Z}]_{:,m},m\in[1,M]$.
Assume that the Gaussian observation of column vector $[\textbf{X}]_{:,m}$ at the $t$-th iteration is given by
\begin{equation}
[\textbf{S}^{t}]_{:,m}=[\textbf{X}]_{:,m}+\bm{n}^t_m,m\in[1,M],
\end{equation}
where $\bm{n}_m^t\sim\mathcal{CN}(\bm{0},v_m^t\textbf{I})$ is the complex Gaussian random vector with variance $v_m^t=\tfrac{1}{2^J}\mathbb{E}\{\|[\textbf{S}^{t}]_{:,m}-[\textbf{X}]_{:,m}\|^2\}$. Starting with $[\textbf{S}^0]_{:,m}=\bm{0}$, $v_m^0=1$ and $t=1$, the linear MMSE (LMMSE) estimation of $[\textbf{X}]_{:,m}$ at the $t$-th iteration in column-by-column way can be computed as
\begin{equation}\label{LMMSE_r}
[\hat{\textbf{R}}^t]_{:,m} = \hat{\textbf{B}}^{t-1}_{m}([\textbf{Y}]_{:,m} -\textbf{C}[\textbf{S}^{t-1}]_{:,m}), m \in[1,M]
\end{equation}
with
\begin{equation}\label{LMMSE_B}
\hat{\textbf{B}}^{t-1}_{m} = v_m^{t-1}\textbf{C}^H\left(v_m^{t-1}\textbf{C}\textbf{C}^H + (\sigma^2)^{t-1}\textbf{I}\right)^{-1}.
\end{equation}

To guarantee the convergence of iterative algorithm, the output of linear estimator $\gamma(\cdot)$ at the $t$-th iteration is imposed a scaling and an orthogonalization on LMMSE estimation as follows ($m \in[1,M]$)
\begin{equation}\label{output_r}
[\textbf{R}^t]_{:,m} = \gamma([\textbf{S}^{t-1}]_{:,m}) =[\textbf{S}^{t-1}]_{:,m} + \frac{2^J}{\textmd{tr}(\hat{\textbf{B}}^{t-1}_{m}\textbf{C})} [\hat{\textbf{R}}^t]_{:,m},
\end{equation}
and the corresponding error variance can be calculated as
\begin{equation}\label{output_u}
u^t_m = v_m^{t-1}[\frac{2^J}{\textmd{tr}(\hat{\textbf{B}}^{t-1}_{m}\textbf{C})} - 1].
\end{equation}

Next, we consider the non-linear estimator $\phi([\textbf{R}^t]_{j,:})$ based on the row-sparse structure of $\textbf{X}$.
Assume that the Gaussian observation of row vector $\bm{x}_j$ at the $t$-th iteration is given by
\begin{equation}
\bm{r}_j^t=[\textbf{R}^t]_{j,:}=\bm{x}_j+\mathbbm{n}_j^t,j\in[1,2^J],
\end{equation}
where $\mathbbm{n}_j^t\!\!\sim\!\!\mathcal{CN}(\bm{0},u^t\textbf{I})$ is the complex Gaussian random vector with variance $u^t=\tfrac{1}{M}\sum\limits_{m=1}^M u^t_m$.
Under this assumption, the MMSE estimations of $\bm{x}_j$ in row-by-row way are generated as
\begin{align}
&\hat{\bm{s}}^t_{j}=\pi^t_{j} \boldsymbol{\lambda}^t_{j},\label{MMSE_s}\\
&\hat{v}^t= \frac{1}{2^J}\sum\limits_{j=1}^{2^J} \left[\pi^t_{j}\left(\rho^t_{j} + \|\boldsymbol{\lambda}^t_{j}\|_2^2\right) - \|\hat{\bm{s}}^t_{j}\|_2^2\right],\label{MMSE_v}
\end{align}
where posterior activity probability $\pi^t_{j}$, Gaussian posterior mean $\boldsymbol{\lambda}^t_{j}$ and variance $\rho^t_{j}$ are calculated as
\begin{align}
&\pi^t_{j} = [(1/\varepsilon^{t-1}_j-1)(1+ g^{t-1}_j/u^t)^M \\\nonumber
 &\ \ \ \ \ \ \ \ \ \ \ \ \ \ \textmd{exp}\left(\frac{ - g^{t-1}_j \|\bm{r}^t_{j}\|_2^2}{u^t(g^{t-1}_j + u^t)}\right) + 1]^{-1},\\
&\boldsymbol{\lambda}^t_{j}=\frac{g^{t-1}_j}{g^{t-1}_j+u^t} \bm{r}^t_{j},\\
&\rho^t_{j}=\frac{g^{t-1}_j u^t}{g^{t-1}_j+u^t}.
\end{align}
After that, similarly, we carry out the scaling and orthogonalization on MMSE estimations at the $t$-th iteration to guarantee the convergence of OAMP detector as follows
\begin{align}
&v^t = \frac{\hat{v}^t u^t}{u^t-\hat{v}^t},\label{output_v}\\
&\bm{s}^t_{j} =[\textbf{S}^t]_{j,:} = \phi(\bm{r}^t_{j})=v^t[\hat{\bm{s}}^t_{j}/\hat{v}^t-\bm{r}^t_{j}/u^t].\label{output_s}
\end{align}
The derivations and detailed explanations of the above content can be found in our previous work \cite{TSP}.
Notice that when the iteration between two estimators converges, the MMSE estimation $\hat{\textbf{S}}^{t_{Last}}=[\hat{\bm{s}}^{t_{Last}}_1,...,\hat{\bm{s}}^{t_{Last}}_{j},...,\hat{\bm{s}}^{t_{Last}}_{2^J}]^T$ before scaling and orthogonalization is the final estimation of $\textbf{X}$, i.e., $\hat{\textbf{X}}=\hat{\textbf{S}}^{t_{Last}}$, where $t_{Last}$ is the index of last iteration.

\subsection{The estimation of unknown priors}
During the above iteration between $\gamma(\cdot)$ and $\phi(\cdot)$, the prior information including noise variance $(\sigma^2)^{t-1}$, codeword activity probability $\varepsilon_j^{t-1}$ and codeword variance $g_j^{t-1}$ are required simultaneously. However, it is difficult for the BS to acquire these information accurately and timely in advance. Therefore, a parameter estimator $\psi(\cdot)$ is needed and combined with $\gamma(\cdot)$ and $\phi(\cdot)$ to realize the codeword detection.

In general, the maximum likelihood (ML) method is often used for parameter estimation \cite{SP}. However, the likelihood function cannot be directly solved due to the existence of hidden variables. Fortunately, the existing powerful EM algorithm is able to convert the ML problem into the maximization problem of its lower bound. In this case, the problem is done through an ``expectation" step of finding the distribution of the hidden variable and a ``maximization" step of maximizing the likelihood function. Inspired by that, we employ an EM algorithm to learn the codeword activity probability $\varepsilon_j$, codeword variance $g_j$ and noise variance $\sigma^2$, denoted by $\boldsymbol{\omega}\triangleq [\sigma^2,\varepsilon_j,g_j]$.

To be specific, we try to obtain the ML estimations of these unknown parameters, i.e., $\hat{\boldsymbol{\omega}}=\textmd{ arg}\max\limits_{\boldsymbol{\omega}}\ln{P(\textbf{Y}|\boldsymbol{\omega})}$. It can be derived through the following two steps:
\begin{itemize}
    \item Expectation Step: Solve the expectation conditioned on $\textbf{Y}$ with parameters $\boldsymbol{\omega}^{t-1}$.
    \begin{align}
        Q(\boldsymbol{\omega},\boldsymbol{\omega}^{t-1})&=\mathbb{E}\{\ln{P(\textbf{Y},\textbf{X}|\boldsymbol{\omega})}|\textbf{Y},\boldsymbol{\omega}^{t-1}\} \\ &= \int_{\textbf{X}}P(\textbf{X}|\textbf{Y},\boldsymbol{\omega}^{t-1})\ln{P(\textbf{Y},\textbf{X}|\boldsymbol{\omega})}.
    \end{align}

    \item Maximization Step: Find the maximum of conditional expectation.
    \begin{align}
        \boldsymbol{\omega}^t={\rm arg}\max\limits_{\boldsymbol{\omega}}Q(\boldsymbol{\omega},\boldsymbol{\omega}^{t-1}).
    \end{align}
\end{itemize}
The joint optimization of $\boldsymbol{\omega}$ is intractable, thus we divide this ML problem into three independent parts, where one parameter is updated at one time while other parameters are fixed.

\emph{EM update of noise variance:}
The elements of AWGN matrix $\textbf{Z}$ follow the i.i.d. Gaussian distribution $P_{\textbf{Z}}(z;\sigma^2)=\mathcal{CN}(z;0,\sigma^2)$.
Due to
\begin{align}
P(\textbf{Y},\textbf{X}|\boldsymbol{\omega})= c_1 P(\textbf{Y}|\textbf{X};\sigma^2)= c_1\prod\limits_{m=1}^M P([\textbf{Y}]_{:,m}|\textbf{C}[\textbf{X}]_{:,m}, \sigma^2)
\end{align}
with $c_1$ being a constant independent of $\sigma^2$, the EM update can be rewritten as
\begin{align}
(\sigma^2)^t&=\textmd{arg}\max\limits_{\sigma^2}\sum\limits_{m=1}^M \mathbb{E}\{\ln{P([\textbf{Y}]_{:,m}|\textbf{C}[\textbf{X}]_{:,m};\sigma^2)}|\textbf{Y},\boldsymbol{\omega}^{t-1}\}\nonumber\\
&=\textmd{arg}\max\limits_{\sigma^2}\sum\limits_{m=1}^M P(\textbf{C}[\textbf{X}]_{:,m}|\textbf{Y},\boldsymbol{\omega}^{t-1})\\
&\ \ \ \ \ \ \ \ \ \ \ \ \ \ \ \ \ \ \ \ \ \ \ \ \ \ \ln{P([\textbf{Y}]_{:,m}|\textbf{C}[\textbf{X}]_{:,m};\sigma^2)}.
\end{align}
Substituting {\small $P([\textbf{Y}]_{:,m}|\textbf{C}[\textbf{X}]_{:,m};\sigma^2)\!\!=\!\!\mathcal{CN}([\textbf{Y}]_{:,m};\textbf{C}[\textbf{X}]_{:,m},\sigma^2\textbf{I})$} into the above equation and zeroing the derivative, we can derive
\begin{small}
\begin{align}
\sum\limits_{j=1}^{2^J} \sum\limits_{m=1}^M \int_X P([\textbf{CX}]_{j,m}|\textbf{Y},\boldsymbol{\omega}^{t-1})\frac{\textmd{ d}}{\textmd{d}\sigma^2}\ln([\textbf{Y}]_{j,m}|[\textbf{CX}]_{j,m};\sigma^2) = 0.
\end{align}
\end{small}
Therefore, we have
\begin{align}\label{EM_sigma}
(\sigma^2)^t&= \frac{1}{M}\sum\limits_{m=1}^M \int_X \|[\textbf{Y}]_{:,m}-\textbf{C}[\textbf{X}]_{:,m}\|_2^2 P(\textbf{C}[\textbf{X}]_{:,m}|\textbf{Y},\boldsymbol{\omega}^{t-1})\nonumber\\
&=\frac{1}{M}\sum\limits_{m=1}^M\left[\frac{1}{n_0}\|[\textbf{Y}]_{:,m}-\textbf{C}[\hat{\textbf{S}}^t]_{:,m}\|_2^2 + \hat{v}^t\right]
\end{align}
with $\hat{\textbf{S}}^t$ and $\hat{v}^t$ being MMSE estimations in the non-linear estimator $\phi(\cdot)$.

\emph{EM update of codeword activity probability:}
Due to
\begin{align}
P(\textbf{Y},\textbf{X}|\boldsymbol{\omega})= c_2 P_{X}(\textbf{X};\varepsilon,g)= c_2\prod\limits_{j=1}^{2^J} P_{X}(\bm{x}_j;\varepsilon_j,g_j)
\end{align}
with $c_2$ being a constant independent of $\varepsilon$, the EM update can be rewritten as
\begin{align}
\varepsilon^t_j &= \textmd{arg} \max\limits_{\varepsilon\in(0,1)} \mathbb{E}\{\ln P(\bm{x}_{j};\varepsilon_j,g^{t-1}_j)|\textbf{Y},\boldsymbol{\omega}^{t-1}\}\nonumber\\
&= \textmd{arg} \max\limits_{\varepsilon\in(0,1)} \int_X P(\bm{x}_{j}|\textbf{Y},\boldsymbol{\omega}^{t-1}) \ln P(\bm{x}_{j};\varepsilon_j,g^{t-1}_j).
\end{align}
Substituting $P_{X}(\bm{x}_{j};\varepsilon_j,g_j)=(1-\varepsilon_j)\delta_0 + \varepsilon_j\mathcal{CN}(\bm{x}_{j};\bm{0},g_j\textbf{I})$ into the above equation and zeroing the derivative, we can derive
\begin{align}
\int_X P(\bm{x}_{j}|\textbf{Y},\boldsymbol{\omega}^{t-1}) \frac{\textmd{d}}{\textmd{d}\varepsilon_j} \ln P(\bm{x}_{j};\varepsilon_j,g^{t-1}_j) = 0.
\end{align}
Hence, we have
\begin{align}\label{EM_epsilon}
\varepsilon^t_j = \pi^t_{j}
\end{align}
with $\pi^t_{j}=P(\bm{x}_{j}|\textbf{Y},\boldsymbol{\omega}^{t-1})$ being the posterior activity probability in the non-linear estimator $\phi(\cdot)$.

\emph{EM update of codeword variance:}
Similarly, the EM update of $g_j$ can be rewritten as
\begin{align}
g^t_j & = \textmd{arg} \max\limits_{g>0} \mathbb{E}\{\ln P(\bm{x}_{j};\varepsilon^{t-1}_j,g_j)|\textbf{Y},\boldsymbol{\omega}^{t-1}\}\nonumber\\
& = \textmd{arg} \max\limits_{g>0} \int_X P(\bm{x}_{j}|\textbf{Y},\boldsymbol{\omega}^{t-1}) \ln P(\bm{x}_{j};\varepsilon^{t-1}_j,g_j).
\end{align}
Substituting $P_{X}(\bm{x}_{j};\varepsilon_j,g_j)=(1-\varepsilon_j)\delta_0 + \varepsilon_j\mathcal{CN}(\bm{x}_{j};\bm{0},g_j\textbf{I})$ into the above equation and zeroing the derivative, we can derive
\begin{align}
\int_X P(\bm{x}_{j}|\textbf{Y},\boldsymbol{\omega}^{t-1}) \frac{\textmd{d}}{\textmd{d}g_j} \ln P(\bm{x}_{j};\varepsilon^{t-1}_j,g_j) = 0.
\end{align}
Thus, we have
\begin{align}\label{EM_g}
g^t_j = \frac{1}{M}\|\boldsymbol{\lambda}^t_{j}\|_2^2+\rho^t_{j}
\end{align}
with $\boldsymbol{\lambda}^t_{j}$ and $\rho^t_{j}$ being the Gaussian posterior mean and variance in the non-linear estimator $\phi(\cdot)$. So far, the parameter estimator $\left[(\sigma^2)^t,\varepsilon_j^t,g^t_j\right]=\psi\left((\sigma^2)^{t-1},\varepsilon_j^{t-1},g^{t-1}_j\right)$ has been designed.

\emph{Remark:} It is worth pointing out that initial values of these unknown parameters are crucial to the above EM updates. Following a previously relevant work \cite{EM}, $\boldsymbol{\omega}^0=\left[(\sigma^2)^0,\varepsilon_j^0,g_j^0\right]$ are set as
\begin{align}\label{parainit1}
&(\sigma^2)^0=\frac{\|\textbf{Y}\|_F^2}{n_0M\left[(\|\textbf{CX}\|_F^2/\|\textbf{Z}\|_F^2)^0+1\right]},\\
&\varepsilon_j^0=\frac{n_0}{2^J}\max\limits_{c_3>0}\frac{1-2^{J+1}\left[(1+c_3^2)\Phi(-c_3)-c_3\varphi(c_3)\right]/n_0} {1+c^2-2\left[(1+c_3^2)\Phi(-c_3)-c_3\varphi(c_3)\right]},\\\label{parainit3}
&g_j^0=\frac{n_0}{2^JM}\left\|[\textbf{C}^H\textbf{Y}]_{j,:}\right\|_2^2,
\end{align}
where $(\|\textbf{CX}\|_F^2/\|\textbf{Z}\|_F^2)^0$ is usually set as 100, $\Phi(\cdot)$ and $\varphi(\cdot)$ denote the cumulative distribution function (CDF) and probability density function (PDF) of standard normal distribution, respectively. The iteration process of state detection and parameter estimation continues until the estimated codeword state matrix converges.

\subsection{The stitching of disordered codewords}
Based on the estimated codeword state matrix $\hat{\textbf{X}}_l$ in the $l$-th sub-slot, we can determine the sets of active codewords present in each sub-slot. However, it is currently unknown which active codewords originate from the same active UE, making it impossible to recover their complete messages. To address this issue, we need to classify the disordered active codewords into distinct classes based on specific characteristics. These classes represent groups of active codewords transmitted by the same device, and the characteristics used for classification are referred to as class labels. Once the active codewords belonging to the same class are mapped back to binary sub-blocks and concatenated in the correct chronological order, we can obtain the original messages from each active UE.

Specifically, we set a flexible decision threshold $\theta_j$ for different codeword index $j$ to obtain the list of active codewords in each sub-slot.
The specific expression of threshold $\theta_j$ is derived in Section \ref{sec:ana}-B.
Denote the list of active codewords of sub-slot $l$ as $\mathcal{L}^{\textmd{ac}}_l=\{j:\|\hat{\bm{x}}_{j,l}\|_2^2>\theta_j\},l\in[1,L]$, where $\hat{\bm{x}}_{j,l}$ is the $j$-th row of estimated codeword state matrix $\hat{\textbf{X}}_l$ and the $j$ in the list is ordered from smallest to largest. In other words, the $j$-th codeword in codebook $\textbf{C}$, for $\|\hat{\bm{x}}_{j,l}\|_2^2>\theta_j$, is the transmitted codeword from one of active UEs.
Meanwhile, let $\hat{K}_a=|\mathcal{L}^{\textmd{ac}}_1|$, we can obtain $\hat{K}_a$ channel vector estimations and channel fading coefficient estimations as follows, $\forall \hat{k}\in[1,\hat{K}_a],j=[\mathcal{L}^{\textmd{ac}}_l]_{\hat{k}}$,
\begin{align}
\hat{\bm{h}}_{\hat{k},l}&=\hat{\bm{x}}_{j,l}=\bm{h}_{\hat{k},l} + \bm{e} \sim \mathcal{CN}\left(\bm{0},\hat{g}_{\hat{k},l}\textbf{I}\right),\\
\hat{g}_{\hat{k},l}&=\tilde{g}_{\hat{k}}+\hat{v}^{t_{Last}},
\end{align}
where the unknown true $\tilde{g}_{\hat{k}}$ needs to be replaced with EM update result $g_{j,l}^{t_{Last}}(j=[\mathcal{L}^{\textmd{ac}}_l]_{\hat{k}})$ and the estimation error $\bm{e}\sim\mathcal{CN}(\bm{0},\hat{v}^{t_{Last}}\textbf{I})$. $g_{j,l}^{t_{Last}}$ and $\hat{v}^{t_{Last}}$ can be fonud in \eqref{EM_g} and \eqref{MMSE_v}. Due to the flexible setting of thresholds in Section \ref{sec:ana}-B, the proposed Bayesian codeword detection can achieve high detection accuracy under favorable conditions, thereby it is assumed that $\hat{K}_a$ is accurately estimated. As a supplement, we conduct experiments to evaluate the influence of inaccurate $\hat{K}_a$ on the algorithm performance in Section \ref{sec:sim}.

With these estimated information, we employ a Bayesian classification approach to stitch the codewords sent from the same active UE and recover its original long message \cite{BC1,BC2}. Based on the intrinsic correlations between the slot-wise channels experienced by a certain active UE, we pick $L$ codewords from $L$ lists of active codewords in sequence for each active UE. Codewords in one list must be selected by different active UEs due to the no-collision transmission.

First, let $\hat{K}_a$ codewords in $\mathcal{L}^{\textmd{ac}}_1$ naturally be $\hat{K}_a$ different classes, denoted by $\{\mathcal{C}_{k'},k'\in[1,\hat{K}_a]\}$. Define the initial labels of these classes, i.e., the common channel statistics of codewords in the same class, as $\{\xi_{k'}=\hat{g}_{k',1},k'\in[1,\hat{K}_a]\}$.
For $l\in[2,L],\hat{k}\in[1,\hat{K}_a]$, the posterior probability of channel vector estimation $\hat{\bm{h}}_{\hat{k},l}$ in each class will be calculated to predict which class the estimated $\hat{\bm{h}}_{\hat{k},l}$ is actually in. Specifically, the $k'$-th class $\mathcal{C}_{k'},k'\in[1,\hat{K}_a]$ with the maximum posterior probability is the class to which $\hat{\bm{h}}_{\hat{k},l}$ belongs and the corresponding codeword index $[\mathcal{L}^{\textmd{ac}}_l]_{\hat{k}}$ is grouped into this class. That is,
\begin{align}\label{cla_result}
[\mathcal{L}^{\textmd{ac}}_l]_{\hat{k}}\in\mathcal{C}_{k'} &= \textmd{ arg}\max\limits_{\mathcal{C}_{k'}}  P(\mathcal{C}_{k'}|\hat{\bm{h}}_{\hat{k},l})\\\nonumber
&=\textmd{ arg}\max\limits_{\mathcal{C}_{k'}}P(\mathcal{C}_{k'})\prod\limits_{m=1}^M P(\hat{h}_{\hat{k},l,m}|\mathcal{C}_{k'}),
\end{align}
where $P(\mathcal{C}_{k'})=\tfrac{1}{\hat{K}_a}$ represents the prior probability of class $\mathcal{C}_{k'}$, $P(\hat{h}_{\hat{k},l,m}|\mathcal{C}_{k'})$ represents the conditional probability and $\hat{h}_{\hat{k},l,m}$ is the $m$-th element of $\hat{\bm{h}}_{\hat{k},l}$. Due to the Gaussian distribution of $\hat{h}_{\hat{k},l,m}$, the probability of observing $\hat{h}_{\hat{k},l,m}$ given the class $\mathcal{C}_{k'}$ can be expressed as
\begin{align}
P(\hat{h}_{\hat{k},l,m}|\mathcal{C}_{k'}) = \frac{1}{\sqrt{2\pi \xi_{k'}}}{\rm exp}(-\frac{\hat{h}^2_{\hat{k},l,m}}{2\xi_{k'}}).
\end{align}
When all indices of active codewords in list $\mathcal{L}^{\textmd{ac}}_l$ have been grouped, each class label $\xi_{k'}$ is updated as
\begin{align}\label{label_update}
\xi_{k'} = \left((l-1)\xi_{k'}+\hat{g}_{\hat{k},l}\right)/l, k'\in[1,\hat{K}_a],[\mathcal{L}^{\textmd{ac}}_l]_{\hat{k}}\in\mathcal{C}_{k'}.
\end{align}

This is one round of codeword classification. The final class labels of this round are assigned to the initial class labels of the next round, and then the above classification process is repeated. The classification result when the class labels converge is the final result.
After that, concatenate $L$ codewords which belong to the same class in chronological order. The long message consisting of $L$ sub-blocks mapped back by these codewords is one of the decoder outputs. Do this for all classes, the transmitted message set $\{\hat{\bm{m}}_{k'}:k'\in[1,\hat{K}_a]\}$ of all active UEs is obtained.
In summary, the proposed Bayesian joint decoding algorithm can be described as Algorithm 1.

\begin{algorithm}
\caption{: Bayesian Joint Decoding}
\label{alg1}
{\bf Input:}
 The codebook matrix $\textbf{C}$, the received signal $\textbf{Y}$.\\
{\bf Output:} The estimated message.
\begin{algorithmic}[1]
\FOR{$l=1:L$}
\STATE{\textbf{Initialize} the maximum number of iterations $T_D, T_S$, $\hat{\textbf{X}}_l^0=\textbf{0}$, $\textbf{S}^0=\textbf{0}$, $v^0=1$, convergence accuracy $\varpi_D, \varpi_S$, and $\boldsymbol{\omega}^0=[(\sigma^2)^0,\varepsilon_j^0,g_j^0]$ are initialized as \eqref{parainit1}-\eqref{parainit3}.}
\FOR{$t=1:T_D$}
\STATE{--- Bayesian Codeword Detection ---}
\STATE{\% Linear Estimator $\gamma(\cdot)$}
\STATE{Compute LMMSE estimations $\hat{\textbf{B}}_m^{t-1}$ and $[\hat{\textbf{R}}^t]_{:,m}$ according to \eqref{LMMSE_B} and \eqref{LMMSE_r};}
\STATE{Obtain orthogonal outputs $[\textbf{R}^t]_{:,m}$ and $u_m^t$ according to \eqref{output_r} and \eqref{output_u};}
\STATE{\% Non-Linear Estimator $\phi(\cdot)$}
\STATE{Set $u^t=\tfrac{1}{M}\sum\nolimits_{m=1}^M u^t_m$ and $\bm{r}_j^t=[\textbf{R}^t]_{j,:}$;}
\STATE{Compute MMSE estimations $\hat{\bm{s}}_j^t$ and $\hat{v}^t$ according to \eqref{MMSE_s} and \eqref{MMSE_v}, and obtain matrix $\hat{\textbf{S}}^t=[\hat{\bm{s}}^t_1,...,\hat{\bm{s}}^t_j,...,\hat{\bm{s}}^t_{2^J}]^T$; }
\STATE{Compute orthogonal outputs $\bm{s}_j^t$ and $v^t$ according to \eqref{output_s} and \eqref{output_v}, and obtain matrix $\textbf{S}^t=[\bm{s}^t_1,...,\bm{s}^t_j,...,\bm{s}^t_{2^J}]^T$;}
\STATE{\% Parameter Estimator $\psi(\cdot)$}
\STATE{Update $(\sigma^2)^t,\varepsilon_j^t,g_j^t$ according to \eqref{EM_sigma}, \eqref{EM_epsilon} and \eqref{EM_g};}

\IF{$\|\hat{\textbf{S}}^t-\hat{\textbf{X}}_l^{t-1}\|_F^2/\|\hat{\textbf{S}}^t\|_F^2<\varpi_D$}
    \STATE{Save $\hat{\textbf{X}}_l=\hat{\textbf{S}}^t$ and $\bm{g}_l=[g_1^t,...,g_j^t,...,g_{2^J}^t]$;}
    \STATE{Break;}
\ELSE
    \STATE{Set $\hat{\textbf{X}}_l^t=\hat{\textbf{S}}^t$;}
\ENDIF
\ENDFOR

\STATE{--- Bayesian Codeword Stitching ---}
\STATE{\% Hard Decision}
\STATE{Compute decision threshold $\theta_j$;}
\STATE{Judge $\mathcal{L}_l^{\textmd{ac}}=\{j:\|\hat{\bm{x}}_{j,l}\|_2^2>\theta_j,j\in[1,2^J]\}$ and let $\hat{K}_a=|\mathcal{L}_1^{\textmd{ac}}|$;}
\STATE{\% Codeword Splicer $\mu(\cdot)$}
\STATE{Save $\hat{\bm{h}}_{\hat{k},l}=\hat{\bm{x}}_{j,l}$ and $\hat{g}_{\hat{k},l}=g^t_{j,l}+\hat{v}^{t},\forall \hat{k}\in[1,\hat{K}_a],j=[\mathcal{L}^{\textmd{ac}}_l]_{\hat{k}}$;}
\ENDFOR

\STATE{Set initial class label $\{\xi_{k'}^0=\hat{g}_{k',1},k'\in[1,\hat{K}_a]\}$;}
\FOR{$t=1:T_S$}
    \STATE{Let $\xi_{k'}^t=\xi_{k'}^{t-1}$;}
    \FOR{$l=2:L$}
    \STATE{Compute $\mathcal{C}_{k'}$ based on \eqref{cla_result} for $\forall \hat{k},k'\in[1,\hat{K}_a]$, judge the codeword index $[\mathcal{L}^{\textmd{ac}}_l]_{\hat{k}}\in\mathcal{C}_{k'}$;}
    \STATE{Update $\xi_{k'}^t$ according to \eqref{label_update};}
    \ENDFOR
    \IF{$\max(|\boldsymbol{\xi}^t-\boldsymbol{\xi}^{t-1}|)<\varpi_S$}
        \STATE{Break;}
    \ENDIF
\ENDFOR

\STATE{Output $\hat{\bm{m}}_{k'}((l-1)J+1:lJ)=\textmd{demap}(j_l)\in \mathbb{B}^{J\times 1},j_l\in\mathcal{C}_{k'}\cap\mathcal{L}^{\textmd{ac}}_l,k'\in[1,\hat{K}_a],l\in[1,L]$.}
\end{algorithmic}
\end{algorithm}

\section{Performance Analysis of Massive Unsourced Random Access}\label{sec:ana}
In this section, the convergence, complexity and error probability of the proposed Bayesian joint decoding-based massive uncoupled unsourced random access scheme are analyzed. Meanwhile, asymptotic analysis in some extreme cases are performed to provide useful insights for the design of massive unsourced random access.
\subsection{Convergence and Decoding Complexity}
First, we discuss the convergence of the proposed decoding algorithm. There are two essential iterations involved.
For the convergence of codeword detection. The orthogonalization of OAMP in \eqref{output_r}, \eqref{output_u}, \eqref{output_v} and \eqref{output_s} are diverge-free and de-correlated operations (please see \cite{OAMP} for proof), which ensures that $\{u^t, v^t\}$ are monotonically decreasing sequences. Besides, they both have lower bound $0$. Combined with the monotone bounded theorem, we know that codeword detection is convergent. The SE fixed point of detector is derived in Appendix \ref{APP:fixp}.
For the convergence of codeword stitching, because the initial class labels of the next round of classification are assigned by the final updated values of current round, the classification result of the next round must be better than that of the current round. After several rounds of classification, class labels will not change. This indicates that the classification result reaches a steady state and thereby the convergence of the proposed codeword stitching is guaranteed.

Then, let us state the complexity of the proposed decoding algorithm.
On the one hand, the computational complexity of codeword detection in each iteration can be expressed as $\mathcal{O}\left((n_0^2 2^J+n_0^3)M\right)$, which is dominated by matrix multiplication and matrix inverse in LMMSE estimation, norm operation in MMSE estimation and matrix-vector multiplication in parameter estimation. If codebook $\textbf{C}$ has special structures, such as Hadamard or Discrete Fourier Transformation (DFT) matrix, the computational complexity will be reduced to $\mathcal{O}\left(J 2^J M\right)$. On the other hand, the computational complexity of codeword stitching in each round depends on the calculation and comparison of posterior probability, that is $\mathcal{O}\left(M(L-1)K_a^2)\right)$. Combined with the fact presented in simulation that the algorithm converges within 15 iterations, we can draw a conclusion that the proposed Bayesian joint decoding algorithm has a relatively low complexity by exploiting the codebook with a particular structure.

\subsection{Error Probability of Decoding}
To characterize the effectiveness of the proposed decoding algorithm, we focus on the error probability of decoding based on \emph{Assumption}. In the following, error probability of codeword detection and codeword stitching will be derived respectively.
\subsubsection{Bayesian codeword detection}
Define the average detection error probability $P_1$ as
\begin{small}
\begin{align}
P_1=\frac{1}{L}\sum\limits_{l=1}^L\left[P^{\textmd{ac}}\frac{1}{2^J}\sum\limits_{j=1}^{2^J}P^{\textmd{md}}_j +(1-P^{\textmd{ac}})\frac{1}{2^J}\sum\limits_{j=1}^{2^J}P^{\textmd{fa}}_j\right],
\end{align}
\end{small}
where $P^{\textmd{ac}}=1-\left(1-1/2^J\right)^{K_a}\approx K_a/2^J$ denotes the overall codeword activity probability. $P^{\textmd{md}}_j=\textmd{Pr}[\hat{\delta}_j=0|\delta_j=1]$ and $P^{\textmd{fa}}_j=\textmd{Pr}[\hat{\delta}_j=1|\delta_j=0]$ are the misdetection probability and the false alarm probability of the $j$-th codeword, respectively, with $\hat{\delta}_j$ and $\delta_j$ representing the estimated and actual activity indicator of the $j$-th codeword. As mentioned above, the value of decision threshold determines the probabilities of misdetection and false alarm. Hence, we obtain these two error probabilities according to their definitions and the expression of hard threshold.

Recall that the input Gaussian observation of non-linear estimator $\phi(\cdot)$ at $t$-th iteration $\bm{r}_j^t = \bm{x}_j +\mathbbm{n}_j^t$, due to $\mathbbm{n}_j^t\sim\mathcal{CN}(\bm{0},u^t\textbf{I})$ and $P_{X}(\bm{x}_{j};\varepsilon_j,g_j)=(1-\varepsilon_j)\delta_0 + \varepsilon_j\mathcal{CN}(\bm{x}_{j};\bm{0},g_j\textbf{I})$,the random variable
\begin{align}
X\triangleq\begin{cases}
    (\bm{r}^t_j)^H\bm{r}^t_j/(2u^t),\quad \varepsilon_j=0,\\
    (\bm{r}^t_j)^H\bm{r}^t_j/(2(u^t+g^{t-1}_j)),\quad \varepsilon_j=1
    \end{cases}
\end{align}
follows the $\mathcal{X}^2$ distribution with $2M$ degree of freedom \cite{GF3}. In the non-linear estimator, the posterior codeword activity probability can be rewritten as
\begin{small}
\begin{align}
\pi^t_{j} &= \left[(1/\varepsilon^{t-1}_j-1)(1+ g^{t-1}_j /u^t)^M e^{\frac{ - g^{t-1}_j \|\bm{r}^t_{j}\|_2^2}{u^t(g^{t-1}_j + u^t)}} + 1\right]^{-1}\nonumber\\
&= \left[(1/\varepsilon^{t-1}_j-1) e^{-M\left( \frac{ g^{t-1}_j \|\bm{r}^t_{j}\|_2^2}{Mu^t(g^{t-1}_j + u^t)} - \log(1+ g^{t-1}_j /u^t) \right)} + 1\right]^{-1}.
\end{align}
\end{small}
Intuitively, whether $\pi^t_{j}$ approaches $0$ or $1$ depends on the relative magnitude of $\frac{ g^{t-1}_j \|\bm{r}^t_{j}\|_2^2}{Mu^t(g^{t-1}_j + u^t)}$ and $\log(1+ g^{t-1}_j /u^t)$. Based on this observation, the codeword activity decision is set as
\begin{align}
\hat{\delta}_j=\begin{cases}0,\quad
\|\bm{r}^t_{j}\|_2^2<\theta_j, \\
1,\quad \|\bm{r}^t_{j}\|_2^2>\theta_j,
\end{cases}
\end{align}
where threshold $\theta_j=Mu^t(g^{t-1}_j + u^t)\log(1+ g^{t-1}_j /u^t)/g^{t-1}_j$. Then, the probabilities of misdetection and false alarm can be computed as
\begin{align}
P^{\rm md}_j&={\rm Pr}[\hat{\delta}_j=0|\delta_j=1]={\rm Pr}\left(X\leq a_j\right)\\\nonumber
&=F_{2M}(a_j)\\\nonumber
&=\frac{\underline{\gamma}(M,a_j/2)}{\Gamma(M)}
\end{align}
and
\begin{align}
P^{\rm fa}_j&={\rm Pr}[\hat{\delta}_j=1|\delta_j=0]={\rm Pr}\left(X\geq b_j\right)\\\nonumber
&=1-F_{2M}(b_j)\\\nonumber
&=\frac{\overline{\gamma}(M,b_j/2)}{\Gamma(M)},
\end{align}
where $a_j=Mu^t(g^{t-1}_j + u^t)\log(1+ g^{t-1}_j /u^t)/(2g^{t-1}_j(u^t+g_j^{t-1}))=Mu^t\log(1+ g^{t-1}_j /u^t)/(2g^{t-1}_j)$ and $b_j=Mu^t(g^{t-1}_j + u^t)\log(1+ g^{t-1}_j /u^t)/(2u^t g^{t-1}_j)=M(g^{t-1}_j + u^t)\log(1+ g^{t-1}_j /u^t)/(2 g^{t-1}_j)$. $F_{2M}(\cdot)$ is the CDF of $\mathcal{X}^2$ distribution with $2M$ degree of freedom. The Gamma function, the lower and upper incomplete Gamma functions are respectively denoted as
\begin{align}
\Gamma(s)&=\int\nolimits_0^\infty x^{s-1}e^{-x}dx,\\\nonumber
\underline{\gamma}(s,x)&=\int\nolimits_x^\infty t^{s-1}e^{-t}dt,\\\nonumber
\overline{\gamma}(s,x)&=\int\nolimits_0^x t^{s-1}e^{-t}dt.
\end{align}
Therefore, the average detection error probability $P_1$ is
\begin{align}
P_1=\frac{K_a}{2^J} \bar{P}^{\rm md} +(1-\frac{K_a}{2^J}) \bar{P}^{\rm fa}
\end{align}
with $\bar{P}^{\rm md}=\frac{1}{2^J}\sum\nolimits_{j=1}^{2^J}P^{\rm md}_j$ and $\bar{P}^{\rm fa}=\frac{1}{2^J}\sum\nolimits_{j=1}^{2^J}P^{\rm fa}_j$.

It is found that $P^{\rm md}_j$ and $P^{\rm fa}_j$ in Bayesian codeword detection have the following relationship
\begin{align}\label{tradeoff}
P^{\rm md}_j+P^{\rm fa}_j=1-\tau(\theta),
\end{align}
where $\tau(\theta)=F_{2M}(b_j)-F_{2M}(a_j)$. Due to $b_j- a_j=\frac{M}{2}\log(1+\frac{g_{j}}{u^t})>0$, it can be seen that $0<\tau(\theta)<1$ based on the monotone increasing CDF of $\mathcal{X}^2$ distribution. Hence, we can balance the misdetection probability and the false alarm probability by adjusting the threshold.

\subsubsection{Bayesian codeword stitching}
At the receiver, we adopt a Bayesian classification approach to realize the codeword stitching. In what follows, from the perspective of classification error rate (CER), we prove that the proposed Bayesian classification approach is optimal.

For any classification rule, when determining data $\bm{h}_{k,l}$ belongs to class $\mathcal{C}_{k'}$, the CER for a single class is defined as the probability that $\bm{h}_{k,l}$ does not belong to $\mathcal{C}_{k'}$, i.e., $P({\rm CER}|\mathcal{C}_{k'})=1-P(\mathcal{C}_{k'})$. It can be observed that the CER now depends solely on the prior probability of class $\mathcal{C}_{k'}$, which can still be reduced. In this case, by considering the effect of observation $\bm{h}_{k,l}$ on classification decisions, the total CER for a class space with $K_a$ classes and $L$ data points in each class can be represented as
\begin{align}
P_{\rm CER}&= \sum\limits_{k'=1}^{K_a}P(\mathcal{C}_{k'})P({\rm CER}|\mathcal{C}_{k'})\nonumber\\
&=\sum\limits_{k'=1}^{K_a}P(\mathcal{C}_{k'})\left(1-1/L\sum\limits_{k\in\mathcal{C}_{k'}}P(\bm{h}_{k,l}|\mathcal{C}_{k'})\right)\nonumber\\
&=\sum\limits_{k'=1}^{K_a}P(\mathcal{C}_{k'})-\sum\limits_{k'=1}^{K_a}\left(P(\mathcal{C}_{k'})/L\sum\limits_{k\in\mathcal{C}_{k'}} P(\bm{h}_{k,l}|\mathcal{C}_{k'})\right)\nonumber\\
&=1-\sum\limits_{k'=1}^{K_a}\left(1/L\sum\limits_{k\in\mathcal{C}_{k'}} P(\mathcal{C}_{k'})P(\bm{h}_{k,l}|\mathcal{C}_{k'})\right),
\end{align}
where $P({\rm CER}|\mathcal{C}_{k'})=1-1/L\sum\nolimits_{k\in\mathcal{C}_{k'}}P(\bm{h}_{k,l}|\mathcal{C}_{k'})$ is the CER for data $\{\bm{h}_{k,l},k\in\mathcal{C}_{k'},l\in[1,L]\}$ in the $k'$-th class. Intuitively, when the sample data $\bm{h}_{k,l}$ is grouped into the class $\mathcal{C}_{k'}$ that maximizes $P(\mathcal{C}_{k'})P(\bm{h}_{k,l}|\mathcal{C}_{k'})$, the total classification correct rate $(1-P_{\rm CER})$ for the class space can reach its maximum
\begin{align}
\sum\limits_{k=1,l=1}^{K_a,L} \max\limits_{k'} P(\mathcal{C}_{k'})P(\bm{h}_{k,l}|\mathcal{C}_{k'})/L.
\end{align}
This is equivalent to maximizing the Bayesian posterior probability $P(\mathcal{C}_{k'}|\bm{h}_{k,l})$. Therfore, it can be proved that the proposed Bayesian classification achieves the minimum CER, i.e.,
\begin{align}
P_{\rm CER}^{\rm min}=1-\sum\limits_{k=1,l=1}^{K_a,L} \max\limits_{k'} P(\mathcal{C}_{k'})P(\bm{h}_{k,l}|\mathcal{C}_{k'})/L.
\end{align}

Note that the class conditional probability $P(\bm{h}_{k,l}|\mathcal{C}_{k'})$ in Bayesian codeword stitching only needs the relatively magnitude to be used to pick out the maximum, and it doesn't matter if the value is greater than $1$. While the class conditional probability should be normalized when calculating the theoretical CER here.

After codeword detection, the codeword splicer receives a correct alternative codeword with probability $1-\bar{P}^{\rm md}$ and a wrong alternative codeword with probability $\bar{P}^{\rm fa}$. Then, the final error probability of Bayesian joint decoding can be cast as
\begin{align}
P_2&=f(1-\bar{P}^{\rm md},\bar{P}^{\rm fa})\nonumber\\
&= \underbrace{(1-\frac{K_a}{2^J}) \bar{P}^{\rm fa}}\limits_{\rm wrong\ codeword}\\\nonumber
&+{\small \underbrace{\frac{K_a}{2^J}(1-\bar{P}^{\rm md})\left(1-\sum\limits_{k=1,l=1}^{k=K_a,l=L} \max\limits_{k'} P(\mathcal{C}_{k'})P(\bm{h}_{k,l}|\mathcal{C}_{k'})/L\right)}\limits_{\rm wrong\ stitching\ of\ correct\ codeword}.}
\end{align}
This means that false alarm of inactive codewords and wrong stitching of detected active codewords cause errors of Bayesian joint decoding\footnotemark[3]. However, it is difficult to observe the effects of key system parameters on decoding performance and draw some deterministic conclusions from this expression directly. Therefore, asymptotic analysis will be performed below to help us understand the theoretical performance.

\footnotetext[3]{As an overall performance index of unsourced random access, $P_2$ is defined from the perspective of ``codeword" and equivalent to the sum of the traditionally used probability of misdetection and the probability of false-alarm of per-user message, which are defined from the perspective of ``message".}

\subsection{Asymptotic Analysis}
In this part, to facilitate the performance analysis of error probability of Bayesian joint decoding, some key parameters in extreme cases will be considered according to the characteristics of massive unsourced random access. For convenience, we first introduce the following lemmas.
\begin{lemma}
Assume that $n_0,2^J\rightarrow\infty$ with a fixed ratio $n_0/2^J$ and $\textbf{C}$ is a right unitarily invariant matrix, the iterative performance of Bayesian codeword detection can be tracked by the following state evolution: Starting with $t=1$ and $v^1=1$,
\begin{align}\label{SE1}
u^t&=\gamma_{\rm SE}(v^t)=v^t[1/\Omega_{\gamma}^t-1],\\
v^{t+1}&=\phi_{\rm SE}(u^t)=u^t[1/\Omega_{\phi}^{t+1}-1]\label{SE2},
\end{align}
where
\begin{align}
\Omega_{\gamma}^t&=\frac{1}{2^J}{\rm tr}\left\{\textbf{C}^H\left(\frac{\sigma^2}{v^t}\textbf{I}+\textbf{CC}^H\right)^{-1}\right\},\\
\Omega_{\phi}^{t+1}&=1-\frac{1}{Mu^t2^J}\sum_j\left\|\hat{\bm{s}}_j^t-\bm{x}_j\right\|_2^2
\end{align}
with $\hat{\bm{s}}_j^t$ being the MMSE estimation in the non-linear estimator $\phi(\cdot)$. Please see our previous work \cite{TSP} for proof.
\end{lemma}

\begin{lemma}
Based on Lemma 1 and $K_a<n_0$, the fixed point of $u^t$ in the above state evolution will converge to
\begin{align}
u^\infty\approx \frac{\sigma^2}{1-\frac{(2^J-n_0)\varepsilon}{(1-\varepsilon)n_0}}.
\end{align}
Please refer to Appendix \ref{APP:fixp} for proof. In the following analysis, $u^t$ in each expression is replaced by the convergent value $u^\infty$.
\end{lemma}

\subsubsection{The influence of the number of BS antennas on Bayesian joint decoding}
Considering that the BS of 6G wireless networks is equipped with a very large antenna array,  this analysis reveals the influence of the number of BS antennas on the error probability of Bayesian joint decoding in the extreme case.

For {\small $\alpha_j=2a_j/M<1$} and {\small $\beta_j=2b_j/M>1$}, the expressions of $P^{\rm md}_j$ and $P^{\rm fa}_j$ can be scaled in $M$ and expanded to \cite{Gammafun}
\begin{small}
\begin{align}\label{Pmd}
P^{\rm md}_j&=\frac{\underline{\gamma}(M,a_j/2)}{\Gamma(M)}\nonumber\\
&=\frac{1}{2}{\rm erfc}\left(-c_j\sqrt{\frac{M}{2}}\right)-\frac{\exp\left(-\frac{1}{2}Mc^2_j\right)}{\sqrt{2\pi M}}\left(\frac{1}{2a_j/M-1}-\frac{1}{c_j}\right)\nonumber\\
&\ \ \ \ \ \ \ \ \ \ \ \ \ -\emph{o}\left(\frac{\exp(-M)}{\sqrt{M}}\right)\nonumber\\
&=-\frac{\exp\left(-\frac{1}{2}Mc^2_j\right)}{2\sqrt{2\pi M}}\left(\frac{1}{\alpha_j-1}+\frac{1}{c_j}\right)+\emph{o}\left(\frac{\exp(-M)}{\sqrt{M}}\right)
\end{align}
\end{small}
and
\begin{small}
\begin{align}\label{Pfa}
P^{\rm fa}_j&=\frac{\overline{\gamma}(M,b_j/2)}{\Gamma(M)}\nonumber\\
&=\frac{1}{2}{\rm erfc}\left(d_j\sqrt{\frac{M}{2}}\right)+\frac{\exp\left(-\frac{1}{2}Md^2_j\right)}{\sqrt{2\pi M}}\left(\frac{1}{2b_j/M-1}-\frac{1}{d_j}\right)\nonumber\\
&\ \ \ \ \ \ \ \ \ \ \ \ \ +\emph{o}\left(\frac{\exp(-M)}{\sqrt{M}}\right)\nonumber\\
&=\frac{\exp\left(-\frac{1}{2}Md^2_j\right)}{2\sqrt{2\pi M}}\left(\frac{1}{\beta_j-1}+\frac{1}{d_j}\right)+\emph{o}\left(\frac{\exp(-M)}{\sqrt{M}}\right),
\end{align}
\end{small}
where the complementary error function
\begin{align}
{\rm erfc}(x)=\frac{\exp(-x^2)}{\sqrt{\pi}x}\left(1+\emph{o}\left(\frac{1}{x^2}\right)\right)
\end{align}
and {\small $c_j=-\sqrt{2(\alpha_j-1-\log(\alpha_j))}$, $d_j=\sqrt{2(\beta_j-1-\log(\beta_j))}$.} Then, we have
\begin{align}
\lim\limits_{M\rightarrow\infty}P^{\rm md}_j&=\lim\limits_{M\rightarrow\infty}-\frac{\exp\left(-\frac{1}{2}Mc^2_j\right)}{2\sqrt{2\pi M}}\left(\frac{1}{\alpha_j-1}+\frac{1}{c_j}\right)\nonumber\\
&\ \ \ \ \ \ \ \ \ \ \ \ \ +\emph{o}\left(\frac{\exp(-M)}{\sqrt{M}}\right)\nonumber\\
&=0
\end{align}
and
\begin{align}
\lim\limits_{M\rightarrow\infty}P^{\rm fa}_j&=\lim\limits_{M\rightarrow\infty}\frac{\exp\left(-\frac{1}{2}Md^2_j\right)}{2\sqrt{2\pi M}}\left(\frac{1}{\beta_j-1}+\frac{1}{d_j}\right)\nonumber\\
&\ \ \ \ \ \ \ \ \ \ \ \ \ +\emph{o}\left(\frac{\exp(-M)}{\sqrt{M}}\right)\nonumber\\
&=0.
\end{align}
Thus, the average detection error probability
\begin{align}
\lim\limits_{M\rightarrow\infty}P_1&=\lim\limits_{M\rightarrow\infty}\left[\frac{K_a}{2^J} \bar{P}^{\rm md} +(1-\frac{K_a}{2^J}) \bar{P}^{\rm fa}\right]=0.
\end{align}

From another point of view, due to the fact that
\begin{align}
\frac{g_j}{u^\infty}>\log(1+\frac{g_j}{u^\infty})>\frac{g_j}{g_j+u^\infty},
\end{align}
$\lim\limits_{M\rightarrow\infty}\|\hat{\bm{x}}_j\|>\theta_j$ when $\delta_j=1$ and $\lim\limits_{M\rightarrow\infty}\|\hat{\bm{x}}_j\|<\theta_j$ when $\delta_j=0$ always hold. In other words, the codeword activity decision will be always right when $M$ is large enough. So $P_1$ goes to zero.

For Bayesian codeword stitching, $M\rightarrow\infty$ means that the number of attribute of data to be classified tends to infinity, which will cause CER to increase. However, $P_1$ goes to zero faster than $P_{\rm CER}$ goes up, so that the final decoding error probability $P_2$ still shows a downward trend as $M\rightarrow\infty$. On the other hand, as analyzed in Section \ref{sec:ana}-A, large $M$ indicates high complexity. Hence, it is necessary to choose an appropriate $M$ to balance the error probability and complexity of decoding.

Additionally, it is interestingly found that due to channel hardening in the scenario with a large-scale antenna array \cite{chanhard}, Bayesian classification in codeword stitching will be degenerated to random classification. In this case, the final error probability of Bayesian joint decoding when $M\rightarrow\infty$ can be simplified to
\begin{align}
\lim\limits_{M\rightarrow\infty}P_2&=(1-\frac{K_a}{2^J})\lim\limits_{M\rightarrow\infty}\bar{P}^{\rm fa}\nonumber\\
&\ \ \ \ \ +\frac{K_a}{2^J}(1-\lim\limits_{M\rightarrow\infty}\bar{P}^{\rm md})\lim\limits_{M\rightarrow\infty}P_{\rm CER}\nonumber\\
&\approx0+\frac{K_a}{2^J}\left(1-\frac{1}{K_a}\right)\nonumber\\
&=\frac{K_a-1}{2^J},
\end{align}
which demonstrates its performance saturation value.
Note that this performance constraint is affected by the Rayleigh fading channel model-based Bayesian classifier, and we can use preprocessing techniques such as dimension reduction or machine learning methods such as support vector machine to improve the performance of the channel features-based codeword concatenation.

\subsubsection{The influence of the transmit power on Bayesian joint decoding}
Similarly, it is helpful to examine the decoding performance with a sufficiently high transmit power, which will imply an upper bound on performance regardless of transmit costs.

When transmit power $P_t\rightarrow\infty$, combined with $\sigma^2=N_0/(n_0P_t)$, we have
\begin{align}
\lim\limits_{P_t\rightarrow\infty}\alpha_j&=\lim\limits_{P_t\rightarrow\infty}2a_j/M\nonumber\\
&=\lim\limits_{u^\infty\rightarrow0}u^\infty\log(1+g_j/u^\infty)/g_j=0
\end{align}
and
\begin{align}
\lim\limits_{P_t\rightarrow\infty}\beta_j&=\lim\limits_{P_t\rightarrow\infty}2b_j/M\nonumber\\
&=\lim\limits_{u^\infty\rightarrow0}(u^\infty+g_j)\log(1+g_j/u^\infty)/g_j=\infty.
\end{align}
Therefore,
\begin{small}
\begin{align}
\lim\limits_{P_t\rightarrow\infty}P^{\rm md}_j&=\lim\limits_{\alpha_j\rightarrow0}-\frac{\exp\left[-M\left(\alpha_j-1-\log(\alpha_j)\right)\right]}{2\sqrt{2\pi M}}\nonumber\\
&\ \ \left(\frac{1}{\alpha_j-1}+\frac{1}{-\sqrt{2(\alpha_j-1-\log(\alpha_j))}}\right)=0
\end{align}
\end{small}
and
\begin{small}
\begin{align}
\lim\limits_{P_t\rightarrow\infty}P^{\rm fa}_j&=\lim\limits_{\beta_j\rightarrow\infty}\frac{\exp\left[-M\left(\beta_j-1-\log(\beta_j)\right)\right]}{2\sqrt{2\pi M}}\nonumber\\
&\ \ \left(\frac{1}{\beta_j-1}+\frac{1}{\sqrt{2(\beta_j-1-\log(\beta_j))}}\right)=0.
\end{align}
\end{small}
That is, the average detection error probability
\begin{align}
\lim\limits_{P_t\rightarrow\infty}P_1&=\lim\limits_{P_t\rightarrow\infty}\left[\frac{K_a}{2^J} \bar{P}^{\rm md} +(1-\frac{K_a}{2^J}) \bar{P}^{\rm fa}\right]=0.
\end{align}
Yet, it is hard to obtain a closed-form expression of final error probability of decoding $\lim\limits_{P_t\rightarrow\infty}P_2$ due to the CER in an uncertain form when $P_t\rightarrow\infty$. In the simulations, the above theoretical results will be verified.

\section{Numerical Results}\label{sec:sim}
In this part, we conduct extensive simulations to evaluate the effectiveness of the proposed Bayesian joint decoding-based massive uncoupled unsourced random access scheme. Unless extra specified, the main simulation parameters are set as: $K_{\textmd{tot}}=500$, $K_a=50$, $M=32$, $b=96$ bits, $J=12$ bits, $L=8$, and $n_0=1024$. In addition, the codebook matrix $\textbf{C}$ is obtained by $n_0$ randomly selected rows of an $2^J$-point DFT matrix. Thereby, $\textbf{CC}^H=2^J/n_0\textbf{I}$.
The noise power is assumed to be $N_0=-110\ \textmd{dBm}$ and the signal-to-noise ratio (SNR) is defined as ${\rm SNR}_k=\frac{P_t\tilde{g}_{k}}{N_0}$. For convenience, set the minimum receive SNR be $15\ {\rm dB}$.
The path loss model of the uplink channel for UE $k$ is $\tilde{g}_k[\textmd{dB}]=-128.1-37.6\log_{10}(d_k)$ \cite{GF3}. $d_k$ is the distance between UE $k$ and the BS, and it is randomly distributed in $(0,0.5]$ km.

\begin{figure}[h] \centering
\includegraphics [width=0.38\textwidth] {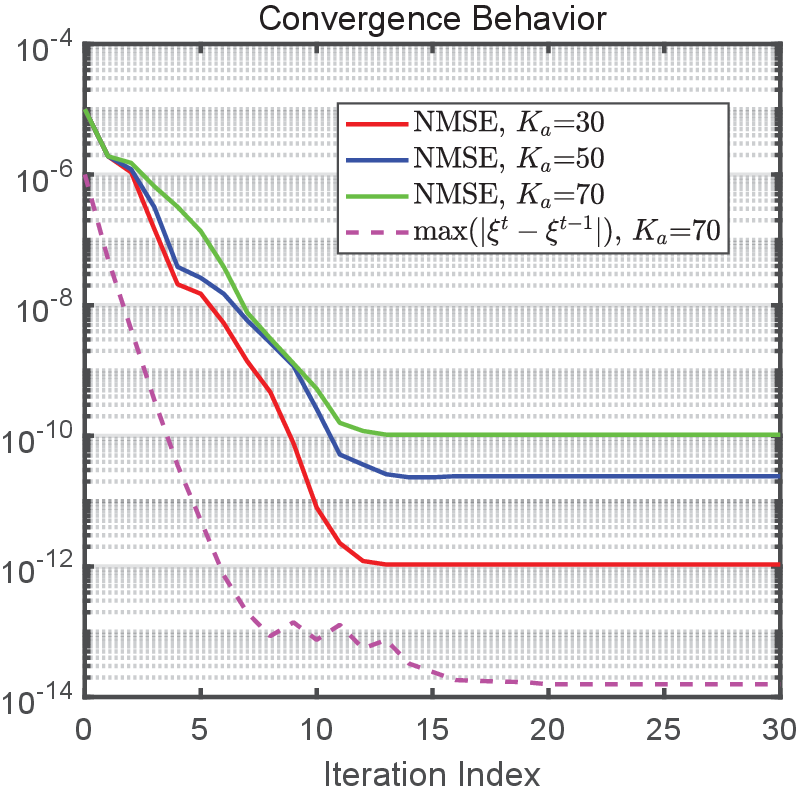}
\caption {The Convergence behaviour for different numbers of active UEs. $J=14$ and $b=112$.} \label{sim0}
\end{figure}
We first observe the convergence of the proposed algorithm with different numbers of active UEs. As shown in Fig. \ref{sim0}, the mean normalized MSE (NMSE) of the proposed codeword detection, i.e., $\textmd{NMSE}=\tfrac{1}{2^JM}\|\hat{\textbf{X}}-\textbf{X}\|_F^2/\|\textbf{X}\|_F^2$, which is presented by three solid lines, converges within 15 iterations with precision $\varpi_D=1e-5$ for all three cases. Additionally, the convergence of class labels, i.e., the convergence of the proposed codeword stitching, which is presented by the dotted line, can also be guaranteed within 15 iterations with precision $\varpi_S=1e-15$. These facts all imply the low complexity of the proposed Bayesian joint decoder.

\begin{figure}[h] \flushright
\includegraphics [width=0.46\textwidth] {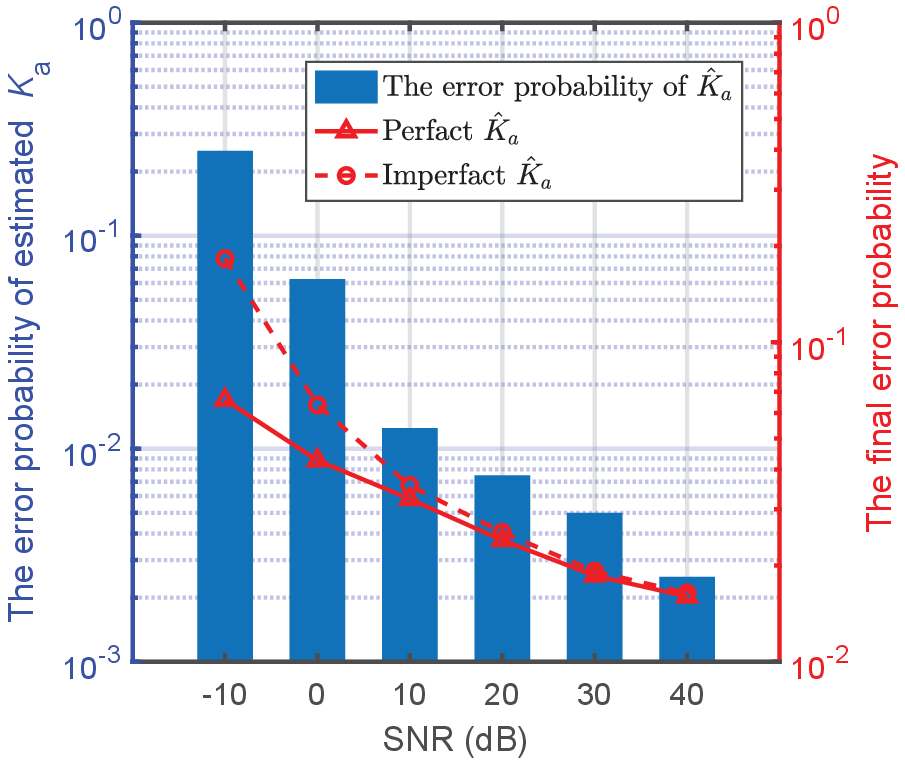}
\caption {The error probability of $\hat{K}_a$ versus SNR and the final error probability $P_2$ versus SNR for perfect and imperfect $\hat{K}_a$.} \label{sim1}
\end{figure}

Fig. \ref{sim1} examines the impact of SNR and the accuracy of $\hat{K}_a$ on the performance of the proposed Bayesian joint decoding algorithm. In Fig. \ref{sim1}, the blue bars represent the error probabilities of $\hat{K}_a$ (the estimated $\hat{K}_a$ is larger or smaller than the true $K_a$) under different SNR levels. The red solid line indicates the scenario where $\hat{K}_a$ is assumed to be perfectly estimated. This line solely reflects the influence of SNR on the final error probability. On the other hand, the red dashed line represents the situation where the BS cannot accurately determine $K_a$ active codewords in each sub-slot after the proposed Bayesian codeword detection. This line demonstrates the combined effect of both SNR and the estimation error of $\hat{K}_a$ on the algorithm's performance.
Intuitively, as the SNR increases, the gap between the two red lines gradually narrows, and they almost coincide when the SNR exceeds 10 dB. This is because, at low SNR levels, the accuracy of estimated $\hat{K}_a$ is relatively poor, which makes the proposed codeword stitching not work well. When the SNR surpasses a certain threshold, the estimation error of the number of active devices becomes negligible in its impact on the performance of the proposed Bayesian joint decoding algorithm. Therefore, under favorable conditions, it is reasonable to assume that $\hat{K}_a$ is accurately estimated.

\begin{figure}[h] \centering
\includegraphics [width=0.4\textwidth] {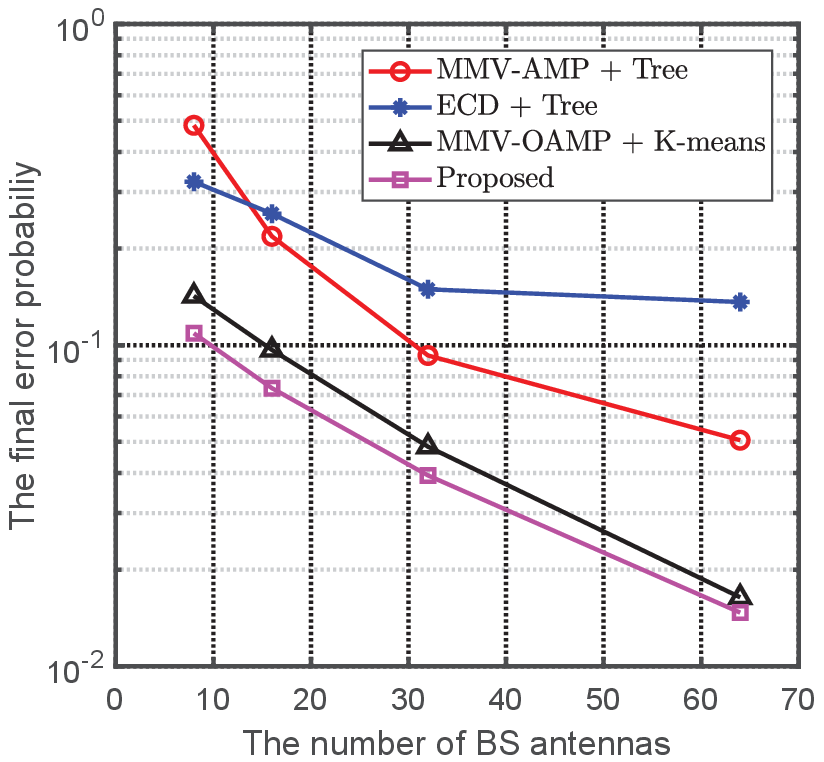}
\caption {The final error probability versus the number of BS antennas for different unsourced random access schemes.} \label{sim12}
\end{figure}
Fig. \ref{sim12} compares the final error probability of the proposed Algorithm 1-based uncoupled unsourced random access scheme with different unsourced random access schemes versus the number of BS antennas.
For comparison, we conduct the following schemes.

(i) Coupled unsourced random access with MMV-AMP codeword detection used and tree code stitching used in \cite{UN5} (written as ``MMV-AMP + Tree" in the legend). To ensure the same length of message $b=96$ bits, the message is divided into $L=32$ sub-blocks of length $J=12$ bits with parity check in $L$ sub-blocks occupying $\{0,9,...,9,12,12,12\}$ bits. For fairness, the same EM parameter estimation as in this paper is employed.

(ii) Coupled unsourced random access with energy codeword detection (ECD) and tree code stitching (written as ``ECD + Tree" in the legend). For codeword detection, compute the codeword energy by correlating the received signal with each column in codebook and obtain the indices corresponding to $K_a$ largest values. For tree code stitching, the parameter settings are the same as scheme (i).

(iii) Uncoupled unsourced random access with MMV-OAMP codeword detection in this paper and K-means stitching adopted in \cite{UN12,UN13} (written as ``MMV-OAMP + K-means" in the legend). The parameters are set the same as this paper.

Intuitively, the error probabilities of all the schemes decrease monotonically as the number of BS antennas increases due to the array gains and the proposed scheme with OAMP detector and Bayesian splicer has lower error probability.
Furthermore, from the perspective of spectral efficiency (SE) per user per channel use, the SE of the proposed scheme and scheme (iii) can be calculated as $SE_0=SE_3=\frac{J}{n_0}=\frac{12}{1024}$ bits/user/channel-use.
Since extra parity check bits introduced in (i) and (ii) occupy some positions, their SE are given by $SE_1=SE_2=\frac{b}{Ln_0}=\frac{8\times12}{32\times 1024}=\frac{1}{4}SE_0$. However, the coupled transmission schemes possess their own advantages. They do not exhibit the error floor effect that arises from channel features-based codeword stitching and have the potential to accommodate a larger number of active devices.
For the proposed scheme and scheme (iii), which have the same SE, it can be seen that when the 3D antenna deployment of the original solution in \cite{UN13} is missing, that is, the additional angle domain information is unavailable, the proposed Bayesian classifier outperforms the K-means method with a small number of antennas. This proves the fact that Bayesian classification has the smallest CER.
Additionally, from the perspective of computational complexity, the complexity of each scheme can be given by

(i) $\mathcal{O}(J2^JM)+\mathcal{O}_{\rm tree}(Ka,L,a_l)$ \cite{UN5},

(ii) $\mathcal{O}(2^Jn_0M^2)+\mathcal{O}_{\rm tree}(Ka,L,a_l)$ \cite{UN5},

(iii) $\mathcal{O}(J2^JM+MLK_a^3)$ \cite{UN13},\\
where the complexity of tree decode $\mathcal{O}_{\rm tree}(Ka,L,a_l)=K_a(L-1)+K_a\sum\nolimits_{n=2}^{L-1}\sum\limits_{m=2}^n K_a^{n-m}(K_a-1)\prod\nolimits_{l=m}^n(2^{-a_l})$ and $a_l$ is the length of parity check in $l$-th sub-block \cite{UN2}.
Obviously, the proposed algorihtm with complexity of $\mathcal{O}(J2^JM+MK_a^2(L-1))$ has lower complexity.
In a nutshell, these facts indicate the feasibility and effectiveness of the proposed algorithm in some scenarios where the receiver requires low computational cost.

\begin{figure}[h] \centering
\includegraphics [width=0.4\textwidth] {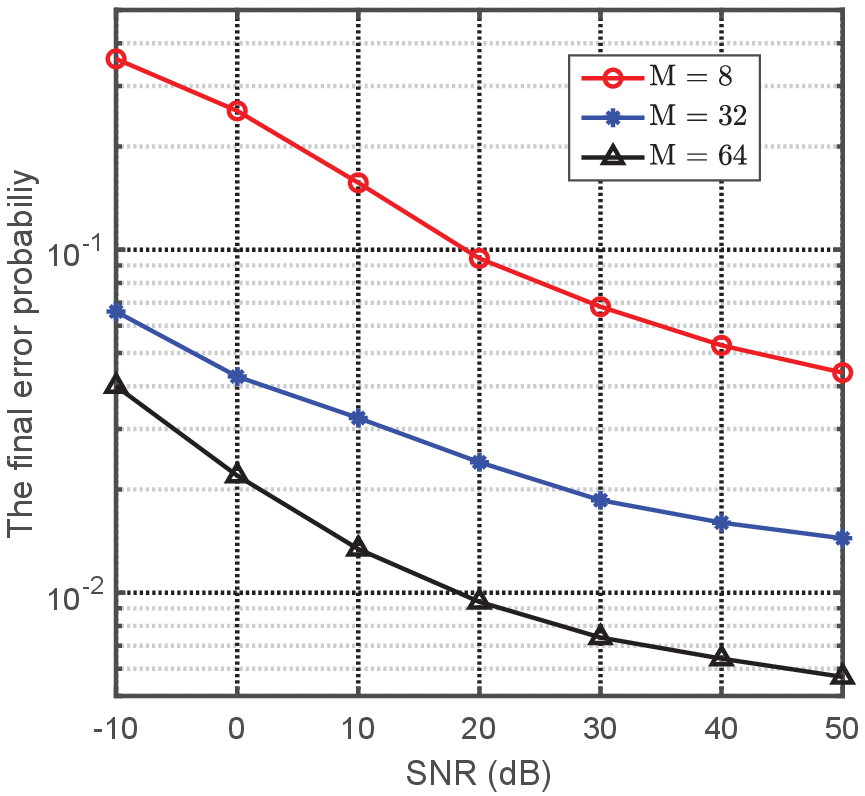}
\caption {The final error probability $P_2$ of the proposed Bayesian joint decoding versus SNR for different numbers of BS antennas.} \label{sim3}
\end{figure}

Then, we show the impact of SNR on the proposed algorithm for different numbers of BS antennas. In Fig. \ref{sim3}, it is seen that the final error probability of Bayesian joint decoding decreases as SNR increases in all cases. Moreover, the decoding performance can be enhanced by adding BS antennas due to the increasing array gains. When the error probability $P_2<0.02$, only $0\ \textmd{dB}$ is needed for $M=64$ while $20\ \textmd{dB}$ is acquired for $M=32$. Thus, it is likely to reduce the transmission costs of device terminals by deploying more antennas at the BS.

\begin{figure}[h] \centering
\includegraphics [width=0.4\textwidth] {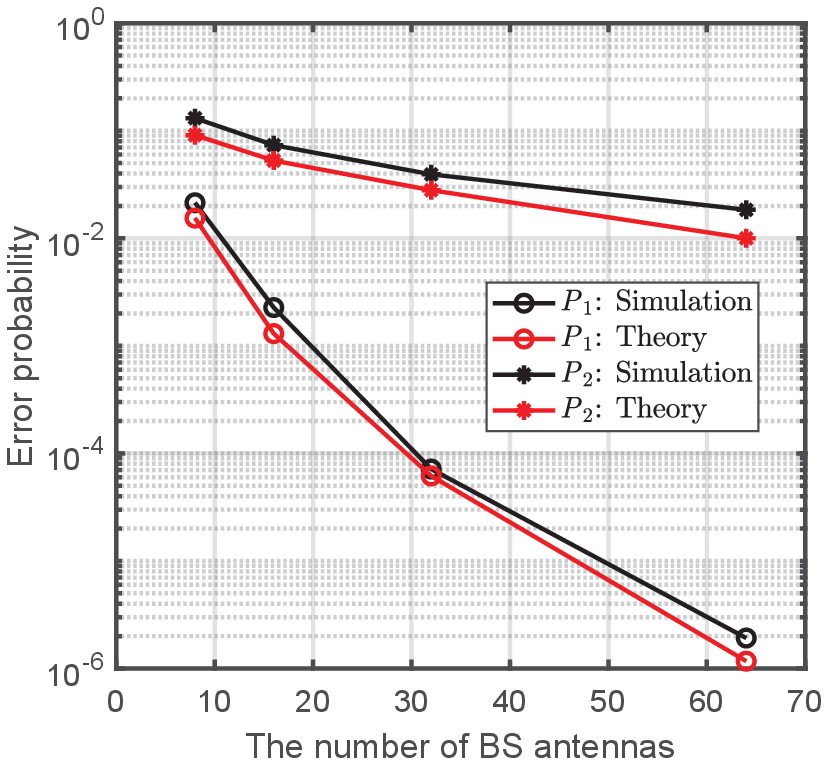}
\caption {The theoretical and simulated error probability versus the number of BS antennas.} \label{sim5}
\end{figure}
Next, we check the asymptotic analysis with the number of BS antennas $M$ in the extreme case. It can be observed from Fig. \ref{sim5} that no matter it is error probability of codeword detection $P_1$ or the final error probability $P_2$, the simulation data obtained from the experiments are basically consistent with the theoretical value calculated from the analysis expression. Specifically, the error probability of codeword detection approaches to zero when the number of BS antennas is sufficiently large, and the final error probability tends to the order of magnitude of $K_a/2^J$.
To explain, since the channel hardening associated with the large number of antennas, the error of Bayesian classification method which employs channel information for codeword stitching tends to be constant in such cases, so that the final error probability decreases with the increment of $M$ at the beginning and tends to saturate when $M$ is large enough.

\begin{figure}[h] \centering
\includegraphics [width=0.4\textwidth] {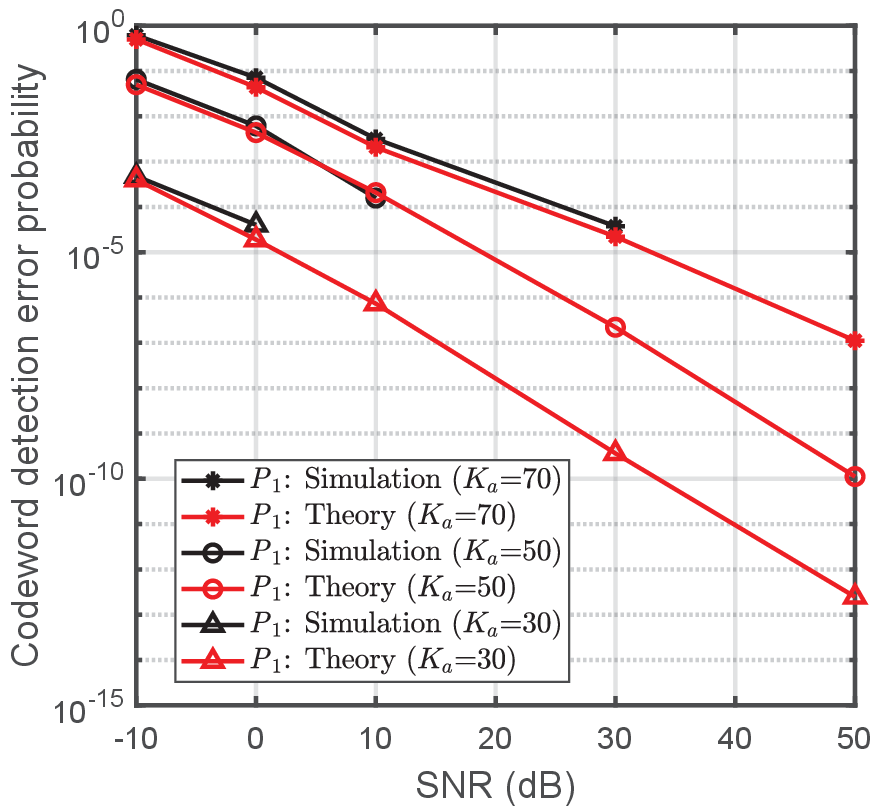}
\caption {The theoretical and simulated codeword detection error probability $P_1$ versus SNR. $M=8$.} \label{sim6}
\end{figure}
Finally, Fig. \ref{sim6} confirms the codeword detection performance when the transmit cost is not a concern. Intuitively, the curves of theoretical and simulated error probability of codeword detection also almost overlap and both of them tend to zero with increment of SNR. When the SNR is large enough, some simulation results are always equal to $0$, such as $10\ \textmd{dB}$ for $K_a=30$, $30\ \textmd{dB}$ for $K_a=50$, and thus cannot be represented in the figure. Meanwhile, it can be found that detection performance improves naturally as the number of active UEs decreases due to less interference.

In summary, the proposed Bayesian joint decoding algorithm has a promising potential of improving the performance of uncoupled unsourced random access in 6G wireless networks.

\section{Conclusion}
This paper proposed a high-efficiency massive uncoupled unsourced random access scheme for 6G wireless networks without requiring extra parity check bits. A low-complexity Bayesian joint decoding algorithm was designed to implement codeword detection and stitching based on channel statistical information. Both theoretical analysis and numerical simulations confirmed that the proposed algorithm had low complexity and good performance. Moreover, asymptotic analysis showed that the error probability of codeword detection tended to zero as the number of BS antennas and the transmit power increased.

\begin{appendices}
\section{Derivation of Lemma 2}\label{APP:fixp}
Based on \eqref{SE1} and \eqref{SE2}, the state evolution with DFT matrix $\textbf{C}$ in Bayesian codeword detection can be simplified as below
\begin{align}
u^t&=\gamma_{\rm SE}(v^t)=\sigma^2+(2^J/n_0-1)v^t,\\
v^{t+1}&=\phi_{\rm SE}(u^t)=\left[1/\Omega_{\rm MMSE}^{t}-1/u^t\right]^{-1},
\end{align}
where
\begin{small}
\begin{align}
\Omega_{\rm MMSE}^{t}&=\frac{1}{M}{\rm E}\left\{\varepsilon_j\left\|\hat{\bm{s}}_j^t-\bm{x}_j\right\|_2^2\right\}\nonumber\\
&=\frac{1}{M}{\rm E}\{\varepsilon_j\}{\rm E}\{\left\|\pi_j\frac{g_j}{g_j+u^t}\left(\bm{x}_j+\mathbbm{n}_j^t\right)-\bm{x}_j\right\|_2^2\}\nonumber\\
&\leq\frac{1}{M}{\rm E}\{\varepsilon_j\}{\rm E}\{\left(\frac{g_j}{g_j+u^t}\right)^2\left(\bm{x}_j+\mathbbm{n}_j^t\right)^H\left(\bm{x}_j+\mathbbm{n}_j^t\right)\nonumber\\
&\ \ \ \ \ \ \ \ \ \ \ \ \ \ \ \ \ \ +\bm{x}_j^H\bm{x}_j -2\frac{g_j}{g_j+u^t}\bm{x}_j^H\bm{x}_j\}\nonumber\\
&=\frac{1}{M}{\rm E}\{\varepsilon_j\}{\rm E}\{\left(1-\frac{g_j}{g_j+u^t}\right)^2\bm{x}_j^H\bm{x}_j\nonumber\\
&\ \ \ \ \ \ \ \ \ \ \ \ \ \ \ \ \ \ +\left(\frac{g_j}{g_j+u^t}\right)^2{\mathbbm{n}_j^t}^H\mathbbm{n}_j^t\}\nonumber\\
&={\rm E}\{\varepsilon_j\}{\rm E}\{\left(1-\frac{g_j}{g_j+u^t}\right)^2g_j+\left(\frac{g_j}{g_j+u^t}\right)^2u^t\}\nonumber\\
&={\rm E}\{\varepsilon_j\}{\rm E}\{\frac{(u^t)^2g_j+g_j^2 u^t}{(g_j+u^t)^2}\}\nonumber\\
&={\rm E}\{\frac{\varepsilon_j g_j u^t}{g_j+u^t}\}.
\end{align}
\end{small}
Thus, we have
\begin{small}
\begin{align}
v^t=\left[1/{\rm E}\left\{\frac{\varepsilon_j g_j u^t}{g_j+u^t}\right\}-1/u^t\right]^{-1}={\rm E}\left\{\frac{\varepsilon_j g_j u^t}{(1-\varepsilon_j)g_j+u^t}\right\}.
\end{align}
\end{small}
Substituting the above equation into the state evolution, the recursive expression can be rewritten as
\begin{align}\label{recur}
u^{t+1}=\sigma^2+(2^J/n_0-1){\rm E}\left\{\frac{\varepsilon_j g_j u^t}{(1-\varepsilon_j)g_j+u^t}\right\}.
\end{align}
Then, we define the function
\begin{align}
f(x)=x-\sigma^2-(2^J/n_0-1){\rm E}\left\{\frac{\varepsilon_j g_j x}{(1-\varepsilon_j)g_j+x}\right\}.
\end{align}
The derivative of $f(x)$ with respect to $x$ is
\begin{small}
\begin{align}
f'(x)=1-(2^J/n_0-1){\rm E}\left\{\frac{\varepsilon_j}{1-\varepsilon_j}\right\}{\rm E}\left\{\frac{g_j^2}{\left(g_j +x/(1-\varepsilon_j)\right)^2}\right\}.
\end{align}
\end{small}
Due to
\begin{align}
0<{\rm E}\left\{\frac{g_j^2}{\left(g_j + x/(1-\varepsilon_j)\right)^2}\right\}<1,
\end{align}
$(2^J/n_0-1){\rm E}\left\{\frac{\varepsilon_j}{1-\varepsilon_j}\right\}<1$, that is $K_a<n_0$, is needed to be satisfied to guarantee $f'(x)>0$, i.e., the monotonicity of $f(x)$. Consequently, the fixed point of state evolution is unique. Next, from \eqref{recur}, we have
\begin{align}
\sigma^2\leq u^t\leq\sigma^2+(2^J/n_0-1){\rm E}\left\{\frac{\varepsilon_j }{1-\varepsilon_j}\right\}u^t,
\end{align}
and it can be further transformed to
\begin{align}
u^t\leq\frac{\sigma^2}{1-\frac{(2^J-n_0)\varepsilon_j}{(1-\varepsilon_j)n_0}}.
\end{align}
Finally, substituting this upper bound of $u^t$ into $f(x)$, we have \eqref{eqf} at the top of the next page.
\begin{figure*}
\begin{align}\label{eqf} f(\frac{\sigma^2}{1-\frac{(2^J-n_0)\varepsilon_j}{(1-\varepsilon_j)n_0}})&=\frac{\sigma^2}{1-\frac{(2^J-n_0)\varepsilon_j}{(1-\varepsilon_j)n_0}}-\sigma^2 -(2^J/n_0-1)\frac{\sigma^2}{1-\frac{(2^J-n_0)\varepsilon_j}{(1-\varepsilon_j)n_0}}{\rm E}\left\{\frac{\varepsilon_j g_j }{(1-\varepsilon_j)g_j+\frac{\sigma^2}{1-\frac{(2^J-n_0)\varepsilon_j}{(1-\varepsilon_j)n_0}}}\right\}\nonumber\\
&=\frac{\sigma^2}{1-\frac{(2^J-n_0)\varepsilon_j}{(1-\varepsilon_j)n_0}}-\sigma^2 -(2^J/n_0-1)\frac{\sigma^2}{1-\frac{(2^J-n_0)\varepsilon_j}{(1-\varepsilon_j)n_0}}{\rm E}\left\{\frac{\varepsilon_j}{1-\varepsilon_j}\right\}\underbrace{{\rm E}\left\{\frac{g_j} {g_j+\frac{\sigma^2}{(1-\varepsilon_j)-\frac{(2^J-n_0)\varepsilon_j}{n_0}}}\right\}}\limits_{\approx1}\nonumber\\
&\approx\frac{\sigma^2}{1-\frac{(2^J-n_0)\varepsilon_j}{(1-\varepsilon_j)n_0}}-\sigma^2 -(2^J/n_0-1)\frac{\sigma^2}{1-\frac{(2^J-n_0)\varepsilon_j}{(1-\varepsilon_j)n_0}} {\rm E}\left\{\frac{\varepsilon_j}{1-\varepsilon_j}\right\}\nonumber\\
&=0.
\end{align}
\end{figure*}
Therefore, it has been proved that the fixed point of state evolution is  $u^\infty\approx\frac{\sigma^2}{1-\frac{(2^J-n_0)\varepsilon_j}{(1-\varepsilon_j)n_0}}$.
\end{appendices}

\end{document}